\documentclass[12pt]{article}
\usepackage[a4paper,margin=1in]{geometry}
\usepackage{graphicx}
\usepackage{amsmath,amssymb}
\usepackage{booktabs}
\usepackage{caption}
\usepackage{float}
\usepackage{rotating}   

\usepackage{authblk}    
\usepackage{orcidlink}  

\usepackage{hyperref}   
\hypersetup{hidelinks}  

\usepackage[backend=bibtex,style=numeric]{biblatex}
\title{\textbf{Enhancing GHSL Population Grids Using Hexagon KH-9 Built-up Data: Refining 1970s Rural and Peri-Urban Distributions in Istanbul}}

\author[1*]{Petrus J. Gerrits \orcidlink{0000-0001-5808-0144}}

\author[2]{Efe Erünal \orcidlink{0000-0002-1391-5643}}

\author[2]{M. Erdem Kabadayı \orcidlink{0000-0003-3206-0190}}

\author[1]{Ana Basiri \orcidlink{0000-0002-2399-1797}}

\author[4, 5]{Elif Sertel \orcidlink{0000-0003-4854-494X}}

\affil[1]{School of Geographical \& Earth Sciences, University of Glasgow, Glasgow, United Kingdom}
\affil[2]{Department of History, Koç University, Istanbul, Turkey}
\affil[4]{Department of Geomatics Engineering, Istanbul Technical University, Istanbul, Turkey}
\affil[5]{Garrick Institute for the Risk Sciences, University of California, Los Angeles (UCLA), Los Angeles, USA}
\affil[*]{Corresponding author: \href{mailto:p.gerrits.1@research.gla.ac.uk}{p.gerrits.1@research.gla.ac.uk}}

\date{} 

\begin{document}

\maketitle

\begin{abstract}
Accurate reconstruction of historical population distributions from the 1970s to the 1990s remains a significant limitation in global gridded population products, owing to coarse built-up data and limited census records. This study is the first, to our knowledge, to integrate declassified Hexagon KH-9 reconnaissance imagery into gridded population mapping. We enhance the GHS-POP framework by combining segmented built-up landcover from the HexaLCSeg dataset (derived from 1977 KH-9 imagery) with geocoded settlement-level census to construct high-resolution historical population grids.

Applied to Arnavutköy and Çekmeköy, Istanbul (1975–1990), we evaluate three dasymetric variants: 1) a standard GHSL baseline, 2) a Hexagon-enhanced workflow, and 3) a fully integrated model with local census records. Pixel-wise and zonal analyses show that GHSL misallocates populations to historically undeveloped regions. On the other hand, our Hexagon-derived dataset greatly improves the representation of fragmented rural and peri-urban areas that are often missing from global products. The integration of settlement-level (LAU-2) census data further refines spatial distribution for our study area.

Results demonstrate that combining historical reconnaissance imagery with high-resolution census data improves the accuracy of historical population grids. Given the global coverage of declassified missions, this methodology offers significant potential for reconstructing historical population patterns in data-scarce regions worldwide.

\end{abstract}

\vspace{0.5em}

\section{Introduction}

Accurate, high-resolution, spatially explicit population data are the foundation for the study of urbanisation dynamics, land-use transitions, disaster risk and demographic change. Global gridded population products such as the Global Human Settlement Layer (GHSL), Gridded Population of the World (GPW), History Database of the Global Environment (HYDE), LandScan and WorldPop have become indispensable for research, policy and assessing progress of Sustainable Development Goals (SDG) \cite{freire_development_2016, calka_ghs-pop_2020, leyk_spatial_2019, european_commission_ghsl_2023, qiu_disaggregating_2022}. These datasets each have different construction methods and scopes (e.g., census-based vs.~modelled, ambient vs.~residential population, etc.), which can lead to substantial differences in their representations of population distribution and overall fitness--for--use \cite{leyk_spatial_2019}. Yet despite their importance, major challenges remain regarding both the spatial and temporal accuracy of these datasets, particularly for periods before the 2000s, as we highlight below.

Recent systematic assessments have revealed notable spatial biases in current global grids: urban populations tend to be better represented than rural ones, while rural populations are often dramatically undercounted \cite{lang-ritter_global_2025}. For instance, Láng-Ritter et al.~\cite{lang-ritter_global_2025} demonstrate that all major global grids significantly underestimate rural populations, with negative biases ranging from approximately $-53\%$ to $-84\%$ when validated against reported figures from large-scale resettlement projects. These shortcomings are largely linked to calibration, resolution, or modelling processes, of which some depend on urban-centric proxy variables (e.g., built-up area masks from moderate-resolution satellite imagery or nighttime lights) that tend to overlook dispersed settlements and hamlets \cite{schug_quantifying_2025}. Furthermore, the limited availability of high-resolution, subnational census data for disaggregation, especially for historical periods, constrains the ability of models to accurately distribute population outside urban extents \cite{freire_development_2016}. As a result, rural and peri-urban communities are often under-represented in gridded datasets, raising concerns for applications that require planning or risk assessment across urban-rural gradients \cite{lang-ritter_global_2025}. This issue can be magnified temporally by historical demographics: according to the latest release of the \textit{United Nations World Urbanization Prospects 2025} \cite{united_nations_world_2025}, in 1975 (the earliest baseline for the GHSL dataset), only approximately $33\%$ of the population lived in cities. Consequently, the reliability of back-casted grids for earlier periods becomes more uncertain, as they are more prone to error in the rural and peri-urban areas where the majority of the human population resided.

Beyond these spatial biases, temporal coverage of population products poses an additional challenge. Both LandScan and WorldPop only provide datasets later than 2000 \cite{lebakula_landscan_2025, leyk_spatial_2019}. While GHSL and the HYDE databases offer gridded population estimates for several historical epochs (GHSL provides 1975, 1990 and 2000 baseline grids, and HYDE spans from 10,000~BCE to present in coarse resolution) \cite{freire_development_2016, klein_goldewijk_anthropogenic_2017}, periods earlier than the 2000s are mostly only available at lower spatial detail or through indirect back-projection methods. For example, HYDE relies on several different historical national population datasets that are downscaled using weight maps \cite{klein_goldewijk_estimating_2001,klein_goldewijk_anthropogenic_2017}, which may not capture localised urban growth or expansion in such detail. The discontinued Global Rural--Urban Mapping Project (GRUMP) provided gridded estimates circa 1990--2000 at 1~km resolution \cite{balk_determining_2006}, but it too was constrained by the coarse resolution of inputs and an urban mask approach that can misallocate population at finer scales. In essence, for fast-growing urban areas such as Istanbul and its outlying districts, as is the case for many fast-urbanising metropolitan areas in the Global South, these data limitations can lead to significant under- or overestimation of local population precisely during periods of growth \cite{chatterjee_growth_2020}. Such uncertainties are consistent with findings by Calka et al.~\cite{calka_ghs-pop_2020}, who observed that the difference between modelled and actual population counts becomes increasingly noticeable as the size of the analysis unit decreases. Without improved historical data, long-term analyses of urban transitions and land-use change starting from the mid-20th century remain uncertain. 

\begin{figure}[H]
    \centering
    \includegraphics[width=1\linewidth]{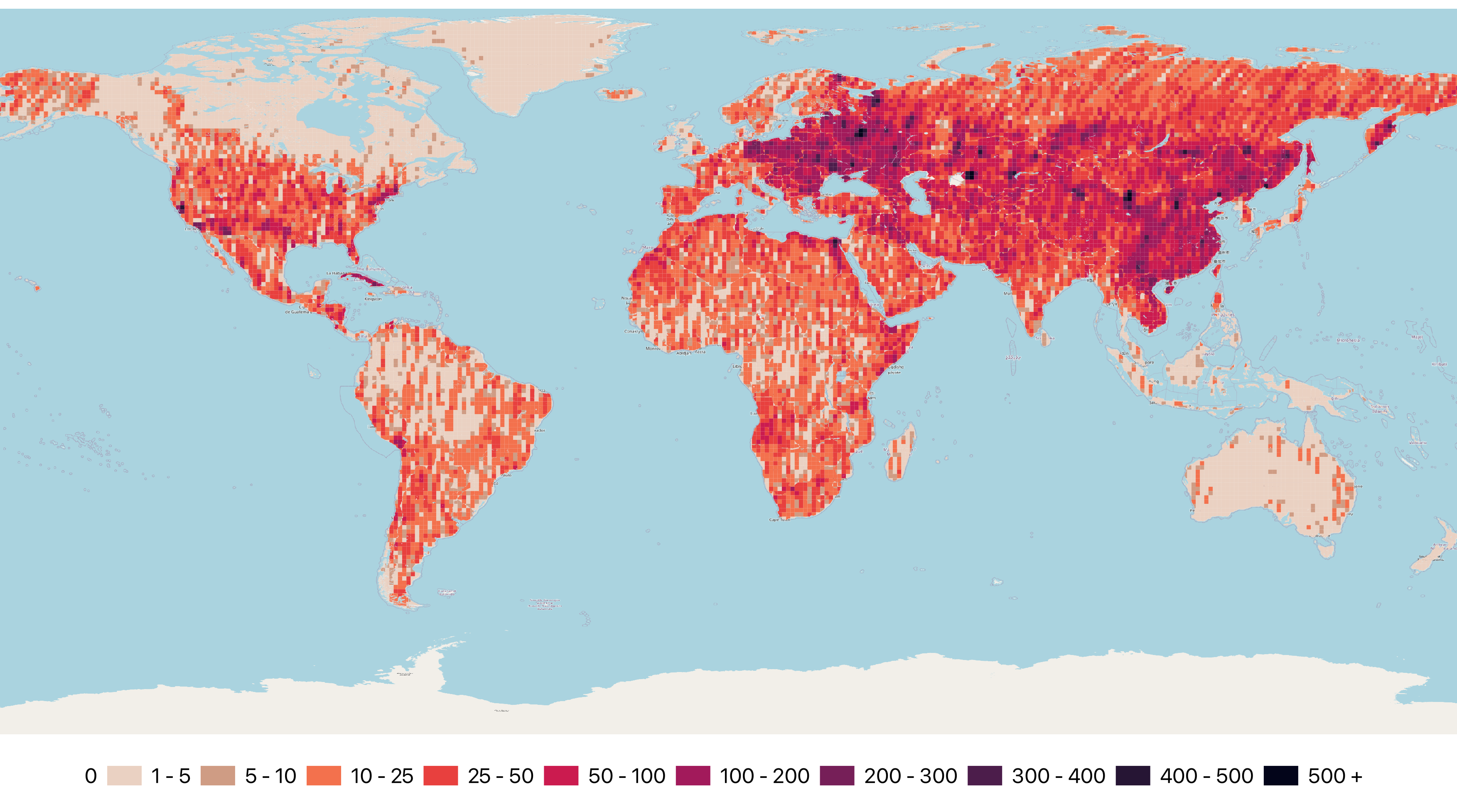}
    \caption{Coverage and count of Hexagon satellite imagery per 1-degree grid cell. Aggregated KH-9 satellite totals were obtained from the USGS website \cite{earth_resources_observation_and_science_eros_center_declassified_2017}. Map data from OpenStreetMap and produced by the authors.}
    \label{fig:hexagon_coverage}
\end{figure}

Innovations in overall compute power and computational methods, remote sensing, and geospatial artificial intelligence (GeoAI) are providing new opportunities to address historical data gaps in a more time-effective manner. More and more high-resolution historical census data are geocoded, and additional historical aerial photographs and satellite imagery are georeferenced and made publicly available. The declassification of Cold War-era reconnaissance imagery, notably the CORONA (KH-4) and HEXAGON (KH-9) satellite programmes, is a particularly interesting satellite dataset \cite{kabadayi_agricultural_2022,earth_resources_observation_and_science_eros_center_declassified_2017,munteanu_potential_2024}. HEXAGON KH-9 imagery (1971--1986) offers globally available panchromatic photographs with ground resolutions on the order of approximately 0.6--1.2~m (2--4~feet). Furthermore, the utilisation of Hexagon imagery in principle is scalable almost globally thanks to the few exceptions (partial coverage of Canada, Greenland, Australia, and Antarctica) to its worldwide coverage; Figure~\ref{fig:hexagon_coverage} shows the global distribution of HEXAGON Declass~3 imagery. This enables the mapping of built-up areas and settlement footprints in the 1960s--1980s with far greater precision than relying on contemporary satellite imagery, as has been shown recently by Sertel et al.~\cite{sertel_hexalcseg_2024} and  by Schug et al.~\cite{schug_quantifying_2025}.  Early applications have demonstrated that KH-9 imagery can support a wide range of quantitative analyses, including elevation change mapping \cite{dehecq_automated_2020}, glacier mass-balance estimation \cite{pieczonka_heterogeneous_2013}, dryland tree monitoring \cite{marzolff_monitoring_2022}, and archaeological landscape reconstruction \cite{hammer_succeeding_2022}, thereby underscoring its suitability for fine-scale historical urban research \cite{shahtahmassebi_establishing_2025}. 

Building on this, subsequent work has shown that integrating HEXAGON and other historic imagery, historical maps and aerial photography can substantially improve the spatial accuracy of reconstructed settlement extents in urban contexts \cite{uhl_combining_2021, uhl_towards_2025}. For instance, Uhl et al.~\cite{uhl_towards_2025} report area-under-curve (AUC) accuracies above 0.93 when using a Grey-Level Co-occurrence Matrix (GLCM) textural measure (PanTex) on historical aerial imagery to delineate 1950s-era settlements, highlighting the potential of such unsupervised proxies to detect potentially sparse rural settlements. Likewise, recent studies have demonstrated that deep learning models can be trained on HEXAGON images to perform land-cover segmentation at very high accuracy (F1 scores around 0.88 for distinguishing built-up land and other classes), effectively creating new ``benchmark'' data for past terrain conditions \cite{hoeser_object_2020, sertel_hexalcseg_2024, schug_quantifying_2025, shahtahmassebi_establishing_2025}. These approaches are increasingly complemented by the digitisation of archival census gazetteers and the adoption of open-source geospatial workflows. Together, they form a methodological toolkit for reconstructing the historical dynamics of population distribution at city and regional scales \cite{uhl_combining_2021,leyk_spatial_2019}. In parallel, ongoing GeoAI developments are driving improved feature extraction and data fusion techniques \cite{huang_advancing_2025,najafabadi_deep_2015}, suggesting that the integration of multi-source historical data can be carried out more systematically and at larger scales than before.

Our motivation and the necessity for our study are best illustrated through a comparison of currently available population products that emphasise the temporal problem. Figure~\ref{fig:pe-comparison} provides the percentage error values for widely used gridded population products relative to district-level census totals in our study area, based on LAU--2 geocoded settlement points (Appendix, table \ref{settlement_table}). While more recent years show improved alignment, errors increase substantially for historical epochs. In particular, the GHS-POP (and GRUMPv1) dataset exhibits consistently elevated percentage errors in the late 1970s and 1980s. This stems from its reliance on back-projection of 2010s statistical records for Türkiye ((For more specific growth rate figures please see the appendix, table \ref{growth_table}). These high percentage errors demonstrate that, precisely in the decades when Istanbul underwent rapid demographic and spatial transformation, global population products diverged most strongly from ground-truth census data. This provides our rationale for the study’s hexagon-enhanced and census-based workflow.

\begin{figure}[H]
\centering
\includegraphics[width=0.8\linewidth]{figures/MAPE_output_results_matrix_2.png}
\caption{Comparative accuracy of global gridded population products for both areas of interest in Istanbul. The heatmaps show the percentage error for GHS-POP, GRUMPv1, LandScan, and WorldPop, computed against official district census totals for benchmark years where each product is available. dark-red cells indicate higher overall percentage error.}
\label{fig:pe-comparison}
\end{figure}

Our study leverages newly digitised and geocoded census records for two Istanbul districts, Arnavutköy and Çekmeköy, together with semantically segmented Hexagon KH-9 imagery from the late 1970s and early 1980s. By adapting a GHSL-style population disaggregation workflow and benchmarking Hexagon-enhanced grids against both census ground truth and existing global datasets (GHSL, GRUMPv1, WorldPop, LandScan), we address two central research questions:
\begin{enumerate}
\item To what extent can the integration of high-resolution legacy Hexagon imagery and harmonised local census data improve the spatial and temporal fidelity of historical population grids in rapidly urbanising peri-urban districts?
\item How do these improved estimates compare with the GHS-POP dataset in capturing the divergent trajectories of urban expansion and rural-urban transformation in Istanbul’s peripheral zones?
\end{enumerate}

Beyond the specific results for Istanbul, this study contributes a reproducible methodological framework for historical population grid generation:
\begin{itemize}
    \item A semi-automated pipeline integrates sub-meter Hexagon imagery into the existing GHS-POP2G workflow to produce the GHS-POP products, providing a quantitative alternative to existing historical products. 
    \item An integrated approach that demonstrates how geocoded census gazetteers can improve population redistribution by capturing intra-district variation overlooked by district-level dasymetric mapping \cite{calka_ghs-pop_2020}, and our experimental design isolates the  contributions of higher-resolution Hexagon KH-9 built-up data and improved lower-level (LAU-2) census-based population counts.

\end{itemize}

\section{Materials and Methods}

\begin{figure}[hb!]
    \centering
    \includegraphics[width=1\linewidth]{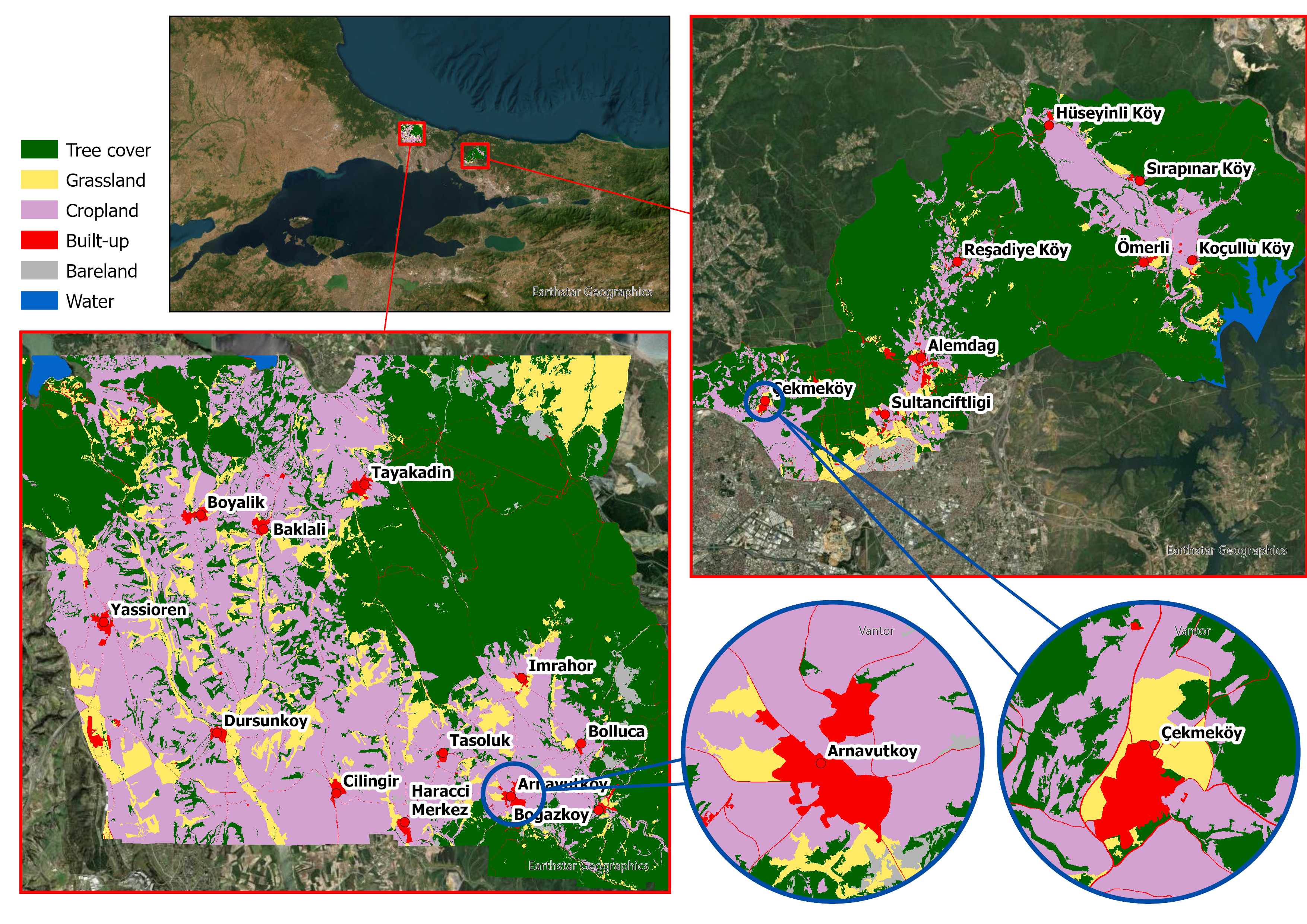}
    \caption{Overview of the study areas: Arnavutköy (west) and Çekmeköy (east), Istanbul. the background classification shows the HexaLCSeg historical landcover (LC) dataset, of which the Built-Up area was derived \cite{sertel_automatic_2024}}
    \label{fig:aoi-overview}
\end{figure}

Our study focuses on two peripheral districts of Istanbul, Arnavutköy and Çekmeköy, which exemplify contrasting trajectories of peri-urban transformation \cite{terzi_urban_2009}. Both were formally established in 2008 through an administrative reorganisation that brought together settlements previously belonging to different jurisdictions to more effectively administer Istanbul’s expanding metropolitan area. These districts were chosen for both their divergent urbanisation dynamics and the availability of detailed census and remote sensing data. Arnavutköy has undergone abrupt land-use change driven by large-scale infrastructure projects, particularly the construction of Istanbul’s new international airport  (construction started in 2015 and expansions are still ongoing), whereas Çekmeköy transitioned more gradually from a rural settlement into a dense suburban district. By including one case of punctuated growth and one of incremental growth, our analysis evaluates how different population disaggregation strategies perform across distinct urban morphologies, a key consideration in recent backcasting and gridded population validation studies \cite{uhl_towards_2025, leyk_spatial_2019}.

\subsection{Study Area}

The Istanbul case study provides a globally relevant test for these questions. Arnavutköy and Çekmeköy exemplify two modes of peri-urban change: abrupt, infrastructure-driven land conversion linked to mega-projects (e.g., new airports and motorways) and gradual, diffuse suburbanisation of rural settlements. Between 1970 and 1990, Istanbul’s population more than doubled, rising from roughly 3 million in 1970 to 5.85 million in 1985 and reaching 7.3 million by 1990. This scale of growth intensified pressure on the rural periphery and accelerated both infrastructural interventions and suburban expansion, with growth spilling into its rural periphery, a pattern that global retrospective datasets have consistently struggled to capture with precision \cite{terzi_urban_2009}. By focusing on these contrasting districts, we evaluate our methods in areas where existing grids are known to be uncertain. More broadly, the study demonstrates that multidisciplinary integration of declassified reconnaissance imagery, digitised census archives, and open geospatial tools can substantially improve the quality and usability of historical gridded population datasets. Although centred on Istanbul, the methods and findings have broader transferability to other rapidly urbanising regions worldwide where historical data are sparse, incomplete, or inconsistently represented.

\begin{figure}
    \centering
    \includegraphics[width=1\linewidth]{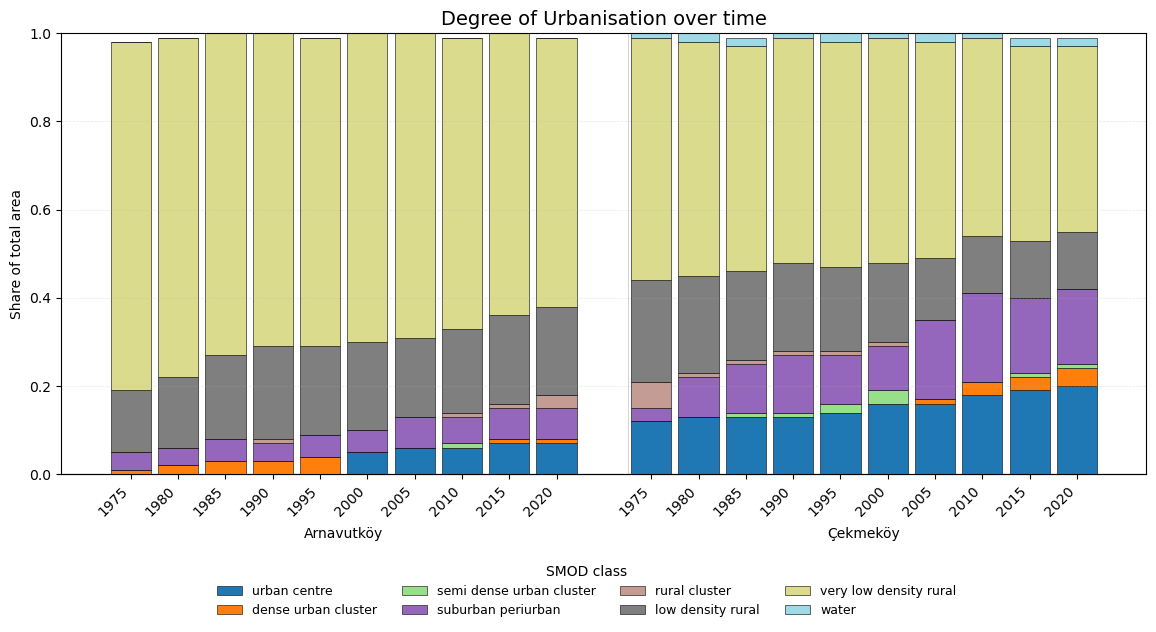}
    \caption{Figure showing the share of area that is attributed to different Degree of Urbanisation classes over time for both the areas of interest according to the
    GHS-SMOD\_GLOBE\_R2023A dataset \cite{schiavina_ghs-pop_2023}}
    \label{smod_aoi_over_time}
\end{figure}

Istanbul itself offers a demanding but globally relevant context for methodological innovation in population estimation. As a megacity, it is characterised by rapid sprawl, diverse peri-urban settlement types, and transformative infrastructure projects within short time spans. These dynamics create a challenging scenario for historical population modelling while also providing insights applicable to other fast-growing global cities where reliable archival data are sparse or inconsistent. Figure \ref{smod_aoi_over_time} illustrates how the Degree of Urbanisation of the study areas has changed over time, from 1975 to 2020 using the GHSL's \textit{Degree of Urbanisation SMOD} dataset. Most notably, we can observe that Arnavutköy increasingly is classified as urban center after 2000, whereas Çekmeköy is already considered an urban center at the baseline of 1975. This is not a trend and classification that we observe from 1977's Hexagon imagery. In section \ref{ref_data} we discuss our selection for the use of this dataset in more detail.

Table~\ref{tab:census_data} presents the historical census totals for Arnavutköy and Çekmeköy from 1935 to 2024. These figures illustrate their divergent demographic trajectories and serve as the ground-truth baseline for our analysis and validation procedures.

\begin{table}[ht!]
\centering
\caption{Historical census population totals for Arnavutköy and Çekmeköy.}
\resizebox{0.5\textwidth}{!}{
\begin{tabular}{lrr}
\toprule
Year & Arnavutköy Census & Çekmeköy Census \\
\midrule
1935 & 5,662 & 2,551 \\
1955 & 9,613 & 6,477 \\
1970 & 8,308 & 3,889 \\
1975 & 8,427 & 8,018 \\
1980 & 11,187 & 10,926 \\
1985 & 14,659 & 16,420 \\
1990 & 42,150 & 34,420 \\
2000 & 96,528 & 88,083 \\
2007 & 125,612 & 139,898 \\
2010 & 145,911 & 152,420 \\
2015 & 181,335 & 208,366 \\
2020 & 227,090 & 248,133 \\
2024 & 266,306 & 281,320 \\
\bottomrule
\end{tabular}}
\label{tab:census_data}
\end{table}

\subsection{Data Acquisition}

To construct and validate historical population grids, we compiled multiple datasets covering census records, global gridded population products, declassified satellite imagery, and ancillary spatial sources. 

\subsubsection{Census Records}

Population counts for all settlements within Arnavutköy and Çekmeköy were obtained from the Turkish Statistical Institute. Census data were available every five years between 1935 and 1990 and then for 2000 as the last census conducted this year. Thereafter the annual figures are based on address-based census compilations. This unique, digitised gazetteer was compiled as part of the UrbanOccupationsOETR \footnote{UrbanOccupationsOETR “Industrialisation and Urban Growth from the mid-nineteenth century Ottoman Empire to Contemporary Turkey in a Comparative Perspective, 1850–2000” project was conducted at Koç University Istanbul, between 2016 and 2022 and was supported by the European Research Council (ERC) under the European Union’s Horizon 2020 research and innovation program Grant Agreement No. 679097.} project, providing georeferenced settlement-level population counts harmonised to 2024 administrative boundaries. These records serve both as the \textit{ground truth} for validation and as inputs for the disaggregation procedure. Their fine spatial granularity allows us to compare district-level aggregates with intra-district settlement variation, an aspect not available in most global population products \cite{leyk_spatial_2019, lang-ritter_global_2025}.

Our settlement-level dataset reflects the demographic pressures and environmental interventions that reshaped the northern and eastern periphery of Istanbul during the late twentieth century. For most census years, the study area consists of 13 settlements in Arnavutköy and 9 in Çekmeköy, yet two exceptions show how the city’s growing population influenced both the physical landscape and the administrative organisation of local settlements. Muratlı in Çekmeköy was submerged following the construction of the Ömerli Dam in 1973, the largest water reservoir supplying Istanbul, which was built in response to rising urban water demand. Although Muratlı remained listed as a legal entity with zero population in the 1975 and 1980 censuses, it disappeared from the records by 1985. In contrast, Nişantepe appeared as a separate populated place for the first time in 2007, reflecting the subdivision of expanding residential areas as the city continued to absorb former rural zones. These two cases illustrate how Istanbul’s demographic expansion produced both a change in the hydrological landscape and a suburban administrative differentiation, creating local discontinuities that complicate the temporal alignment of census data at the settlement-level with historical built-up extents.

\subsubsection{Global Population Grids}

To contextualise the experimental results, several widely used global gridded population datasets were reviewed. Among these, the \textit{Global Human Settlement Layer} (\textit{GHSL}) represents the most directly comparable reference framework. The 2023a release of GHS-POP provides multi-temporal population estimates at 100~m resolution for epochs centred on 1975, 1980, 1985, and 1990 \cite{schiavina_ghs-pop_2023}. Notably, it remains the only global dataset that explicitly integrates Landsat-derived built-up masks for these historical periods and goes back to 1975, roughly corresponding to our KH-9 satellite dataset, making it a particularly suitable baseline for evaluating the Hexagon-enhanced population grids developed in this study.

In contrast, the HYDE database offers long-term reconstructions of global population trends extending back to the early twentieth century \cite{klein_goldewijk_estimating_2001, klein_goldewijk_anthropogenic_2017}. Although HYDE provides valuable contextual information on macro-scale demographic change, its coarse 5 arc min spatial resolution and reliance on aggregated historical statistics limit its applicability for late 20th-century district-level validation within the Istanbul metropolitan region.

More recent population grids, such as \textit{WorldPop} and \textit{LandScan}, achieve higher spatial resolutions (10~m–1~km) but are restricted to post-2000 epochs. These datasets were therefore included only for later qualitative assessment and growth patterns, as their temporal coverage does not extend to the 1975–1990 period analysed here.

Accordingly, the GHSL dataset serves as the primary reference for validation and comparison. Its methodological structure, based on Landsat-derived built-up extents and dasymetric population redistribution—closely parallels the methodology adopted in this research, thereby providing a coherent and consistent benchmark for assessing the accuracy and representativeness of the hexagon-enhanced population grids. Moreover,  GHSL's focus on reproducible characteristics through an open and consistent methodology, designed specifically to support policy-making, makes this dataset the most suitable dataset to analyse in this study \cite{carioli_ghs-pop_2023}.  

\subsubsection{Hexagon KH-9 Imagery}

\begin{figure}[H]
    \centering
    \includegraphics[width=1\linewidth]{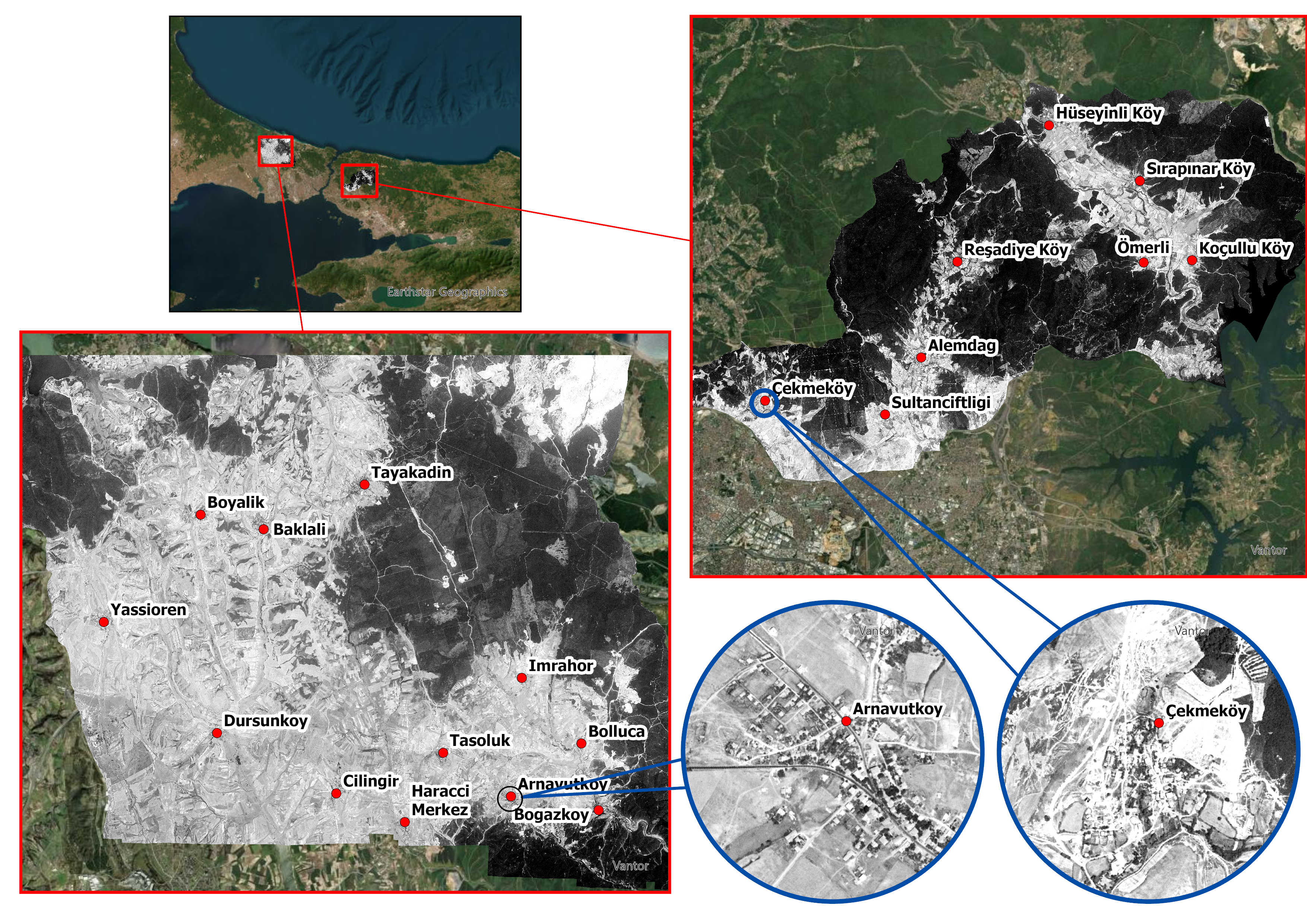}
    \caption{Example of the resolution of Hexagon imagery  for the study areas.}
    \label{fig:hexagon_resolution_example}
\end{figure}

Declassified U.S. National Reconnaissance Office Hexagon KH-9 imagery was obtained for Istanbul in 1977. Figure~\ref{fig:hexagon_resolution_example} With sub-metre panchromatic resolution, the KH-9 series offers historical coverage for reconstructing built-up surfaces in the late 20th century. The imagery was acquired through the U.S. Geological Survey (USGS) archive and scanned at high quality for geometric correction and segmentation as a part of the GeoAI\_LULC\_Seg\footnote{GeoAI\_LULC\_Seg "A GeoAI-based Land Use Land Cover Segmentation Process to Analyse and Predict Rural Depopulation, Agricultural Land Abandonment, and Deforestation in Bulgaria and Turkey, 1940-2040" project was conducted at Koç University Istanbul, between 2022 and 2024 and was supported by the European Research Council (ERC) under the European Union’s Horizon 2020 research and innovation program Grant Agreement No. 101100837.} project.  Compared to Landsat (60~m resolution), Hexagon enables the identification of dispersed rural settlements and small peri-urban clusters that are otherwise omitted in GHSL-derived built-up masks \cite{sertel_hexalcseg_2024}.

\subsection{Data Processing}

All datasets were pre-processed to ensure consistency across space, time, and thematic content. The workflow emphasised three main aspects: reprojection into a common spatial framework, the preparation of Hexagon imagery for built-up surface extraction, and the harmonisation of census data with administrative boundaries.

\subsubsection{Projection and Spatial Framework}

To maintain comparability with the GHSL datasets, all raster and vector data were projected into the World Mollweide projection (EPSG:54009). The Mollweide equal-area system is the standard reference for GHSL \cite{pesaresi_advances_2024}, ensuring that population counts are not biased by distortions in cell size across latitudes. Grids were produced at a target resolution of 100~m, consistent with both GHSL recommendations and the minimum mapping unit detectable from Hexagon imagery.

\subsubsection{Hexagon Imagery Correction and Preparation}

The 1977 KH-9 Hexagon imagery required geometric correction to align with modern spatial references. Kabadayi et al. applied an automated tie-point detection approach supplemented by manual ground control points (GCPs) derived from identifiable stable features, including road intersections and river bends visible in both historical and modern basemaps. Rubber sheeting adjustments were implemented to minimise local distortions, following protocols for legacy aerial and reconnaissance imagery. Radiometric corrections were applied to normalise illumination differences across frames. Following this, the imagery was mosaicked and clipped to the study area, ensuring complete coverage of Arnavutköy and Çekmeköy \cite{kabadayi_analyzing_2024}.

\subsubsection{Semantic Segmentation of Built-up Surfaces}

Hexagon imagery was segmented to derive high--resolution built-up masks for the region using a geographic object-based image analysis (GeOBIA) and the eCognition program (Trimble Germany GmbH, Munich, Germany), following the methodology applied by Sertel et al.~\cite{sertel_hexalcseg_2024, kabadayi_analyzing_2024}. The land cover classes were derived from the ESA land cover classification scheme, including cropland, built-up, bare/sparse vegetation, grassland, shrubland, water bodies, and tree cover (see figure \ref{fig:aoi-overview}). The explanation of each land cover class can be found in “WorldCover\_PUM\_v1.0” \url{(https://worldcover2020.esa.int/}. For the creation of the enhanced GHSL gridded products, we only made use of the built-up classification of the segmented images. Furthermore, Hexagon image selection was carried out through visual inspection and comparison with reference data, confirming that the Hexagon mask captured dispersed rural settlements and peri-urban clusters \cite{sertel_hexalcseg_2024}. 

\subsubsection{Census Data Harmonisation}

Settlement-level census records from the Turkish Statistical Institute  were harmonised to the 2024 locations to ensure continuity. Having obtained the higher level (LAU-2) settlements for our study area,  avoided the boundary-mismatch problem frequently encountered in gridded population studies \cite{leyk_spatial_2019}. Population totals were aggregated to the Areas of Interest and district levels (Arnavutköy and Çekmeköy) and prepared as input for the population disaggregation procedure at a later step.

\subsection{Analysis}

\begin{figure}
    \centering
    \includegraphics[width=1\linewidth]{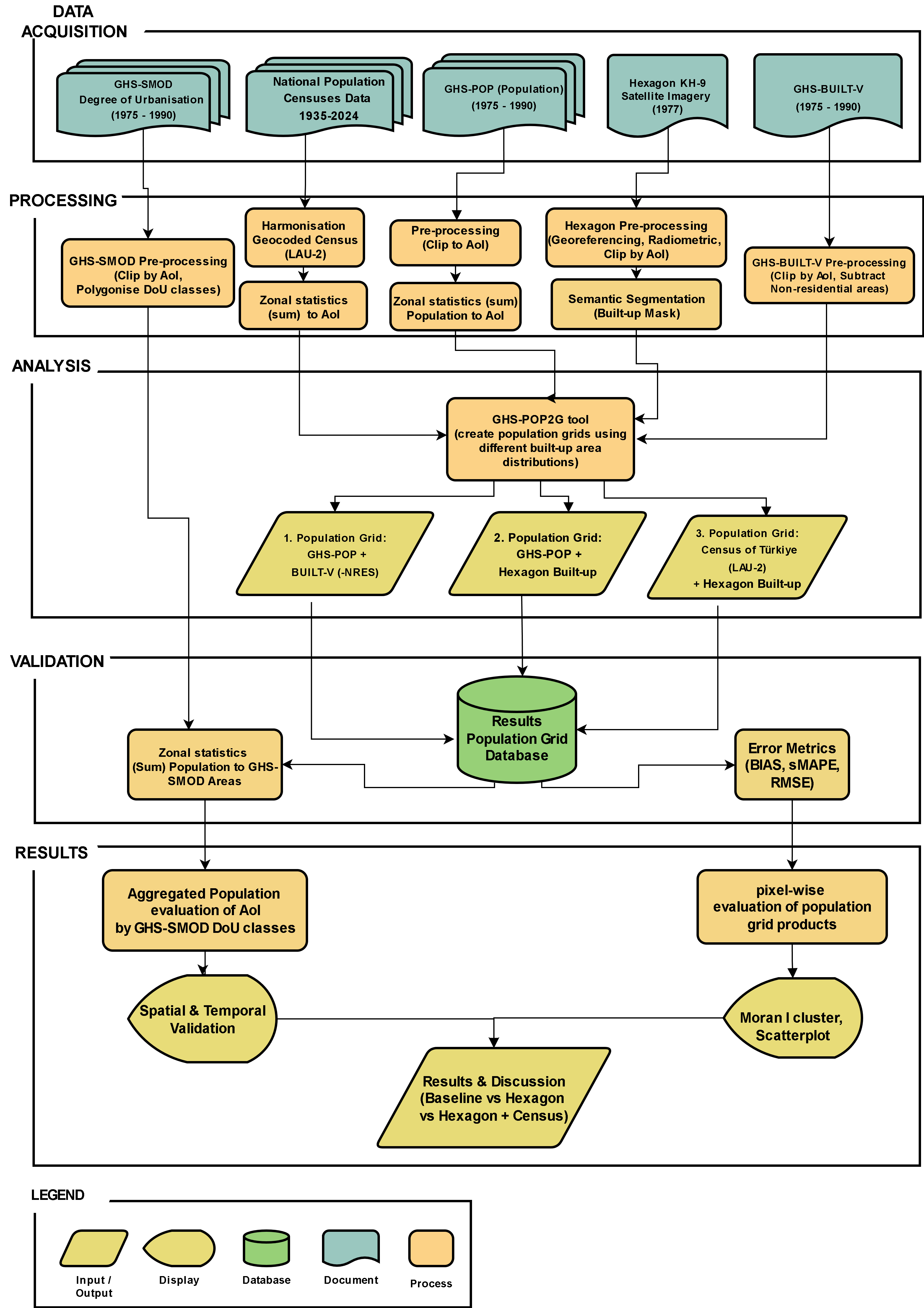}
    \caption{Workflow showing the methods applied in this research.}
    \label{fig:workflow}
\end{figure}

The central aim of this study was to generate historical population grids for Istanbul’s peri-urban districts by adapting and extending the Global Human Settlement Layer (GHSL) \textit{Pop2Grid} methodology \cite{pesaresi_operating_2016, pesaresi_advances_2024}. The general overview of this workflow is presented in figure \ref{fig:workflow}. Alternative population disaggregation was conducted at 100~m resolution using the World Mollweide projection (EPSG:54009), ensuring consistency with GHSL conventions.

\subsubsection{Experimental Grid Products}

We designed a tiered experimental framework, generating and comparing three alternative gridded population products for benchmark years 1975, 1980, 1985, and 1990:
\begin{enumerate}
    \item \textbf{Baseline (GHSL-style)}: Population totals from the Turkish Statistical Institute were disaggregated using Landsat-derived GHSL built-up masks as dasymetric weights.
    \item \textbf{Hexagon-enhanced}: The same census totals were disaggregated using the Hexagon-derived built-up mask in place of the Landsat-based surface, isolating the effect of improved spatial detail in the input imagery.
    \item \textbf{Hexagon + local census}: In addition to the Hexagon mask, this workflow incorporated geocoded settlement-level census data, allowing population allocation to reflect internal variation within districts rather than assuming uniform distribution.
\end{enumerate}
This design allowed us to test separately (i) the effect of higher-resolution built-up mapping (Baseline vs.~Hexagon-enhanced) and (ii) the added value of localised census data (Hexagon-enhanced vs.~Hexagon + local census), while also evaluating the overall gains from both improvements combined (Baseline vs.~Hexagon + local census).

\subsubsection{Disaggregation Procedure}

Population disaggregation was carried out using the GHS-POP2G tool and automated using ArcPy to calculate for multiple years and types \cite{european_commission_joint_research_centre_ghs-pop2g_2023}. The procedure allocated census population totals proportionally across built-up cells within each of our areas of interest. Built-up surfaces (GHS-BUILT or Hexagon) acted as weighting layers, ensuring that population was only assigned to cells classified as built.

For the Hexagon + local census workflow, settlement-level census totals were first spatially matched to their geocoded coordinates and aggregated to the built-up mask. This allowed finer-grained allocation, particularly in districts with dispersed rural populations where uniform disaggregation would misrepresent settlement patterns.

\subsubsection{Consistency Checks}

At each stage, outputs were validated to ensure that district-level totals matched the Turkish Statistical Institute  census reference values. Grids were also visually inspected to confirm realistic allocation of population to settlement structures. Discrepancies identified in preliminary tests (e.g., underestimation of peri-urban growth in Arnavutköy) were traced back to limitations of Landsat-derived masks, underscoring the rationale for Hexagon integration.

Through this tiered and systematic approach, we established a set of gridded population products that could be consistently compared against each other and against external benchmarks, enabling robust evaluation of spatial and temporal fidelity.

\subsection{Validation}

Validation was designed to assess both the aggregate accuracy and the spatial realism of the generated population grids, with particular emphasis on detecting systematic biases and evaluating temporal consistency. 

\subsubsection{Reference Data}
\label{ref_data}

The validation relied on official census totals from the Turkish Statistical Institute, harmonised across administrative changes and available at both district and settlement (LAU-2) levels for 1975, 1980, 1985, and 1990. These figures were used as the authoritative reference for all accuracy assessments (Appendix, table \ref{settlement_table}). 

\begin{figure}
    \centering
    \includegraphics[width=1\linewidth]{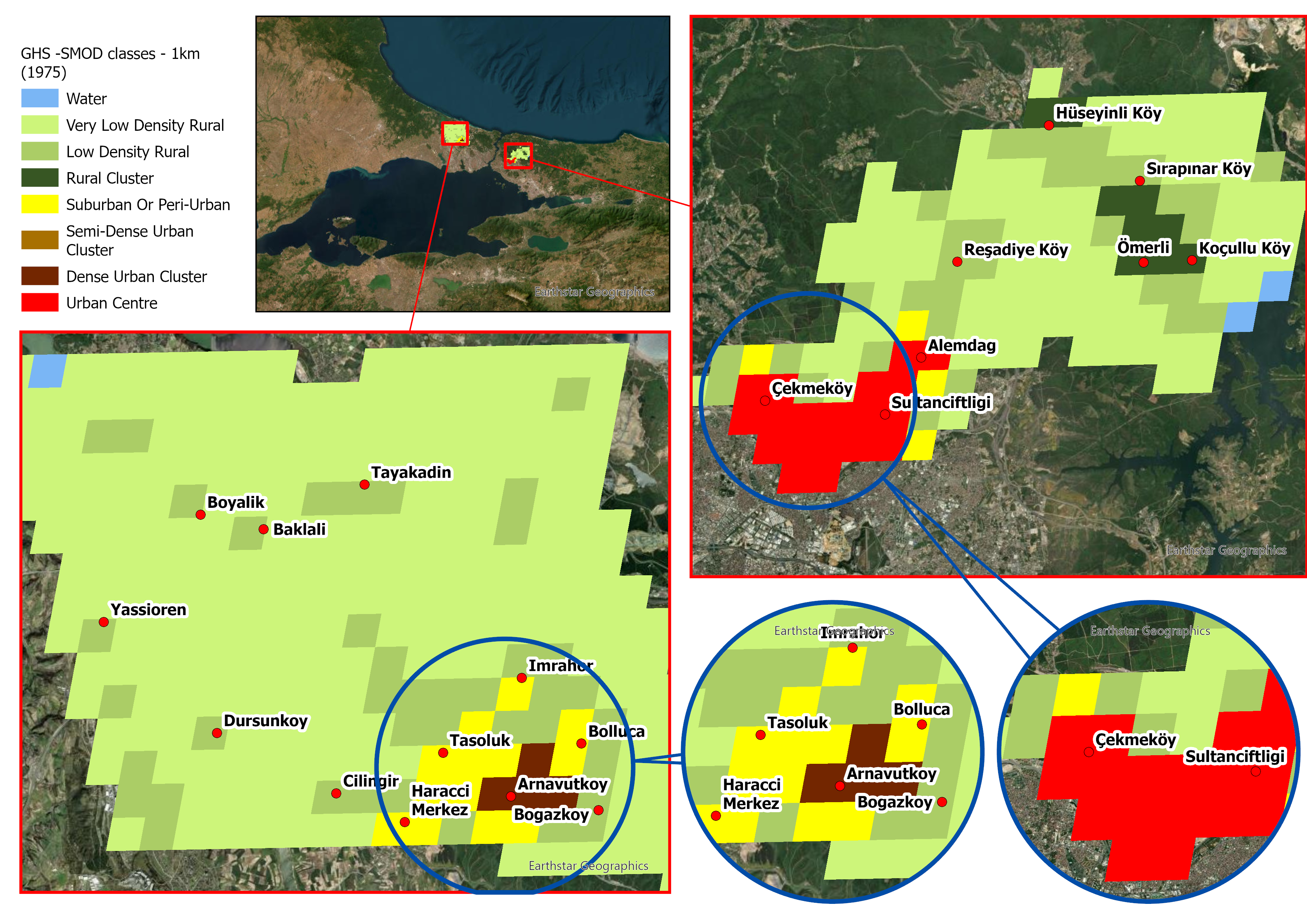}
    \caption{Degree of Urbanisation classification (GHS-SMOD) for the study districts of Çekmekoy and Arnavutköy in 1975. The map reveals that Çekmekoy is classified as an "urban centre" in the baseline year, a designation that appears inconsistent with its growth trajectory against historical census and Hexagon-derived settlement data presented in the results section.}
    \label{smod_overview_1975}
\end{figure}

In addition, the \textit{GHS-SMOD} Degree of Urbanisation (DoU) grid served as the functional zoning framework. GHS-SMOD R2023A employs the Degree of Urbanisation categorisation framework on gridded population and built-up surface data at a 1-km resolution, covering multiple time periods from 1975 to 2030. The product categorises settlements into hierarchical classifications, from cities and towns to villages and rural regions, by integrating population density and built-up area thresholds obtained from the GHS-POP and GHS-BUILT-S datasets. Figure \ref{smod_overview_1975} provides an overview of the degree of urbanisation of the study area 1975. This spatially explicit settlement modelling method allows for gridded examination of urbanisation trends and settlement hierarchy dynamics that are not bound by the areas of interest or administrative boundaries, making it suitable for small area urban-rural evaluations and regional development research \cite{kuffer_missing_2022, schiavina_ghs-pop_2023}. 

\subsubsection{Evaluation Metrics}

Three complementary metrics were employed to evaluate population grid accuracy:

\begin{itemize}
    \item \textbf{Bias (\%)} — measures systematic over- or underestimation of population totals relative to the census reference:
    $$
    \mathrm{Bias} = 
    \frac{\sum P_{\text{predicted},i} - \sum P_{\text{reported},i}}
         {\sum P_{\text{reported},i}} \times 100
    $$
    where positive values indicate overestimation and negative values underestimation of population counts. A bias of 0\% suggests perfect aggregate agreement.

    \item \textbf{Symmetric Mean Absolute Percentage Error (sMAPE)} — quantifies the average proportional deviation between predicted and reported populations, normalised by their combined magnitude:
    $$
    \mathrm{sMAPE} = 
    \frac{1}{n}\sum_{i=1}^{n} 
    \frac{|P_{\text{reported},i} - P_{\text{predicted},i}|}
    {|P_{\text{reported},i}| + |P_{\text{predicted},i}|}
    $$
    sMAPE values range from 0 (perfect fit) to 1 (100\% mean deviation). This metric is less sensitive to scale and therefore more suited for rural and peri-urban domains where population totals vary widely, as other studies have suggested \cite{lang-ritter_global_2025}.

    \item \textbf{Root Mean Square Error (RMSE)} — captures absolute deviations between predicted and reported populations, expressed as:
    $$
    \mathrm{RMSE} = 
    \sqrt{\frac{1}{n}\sum_{i=1}^{n}(P_{\text{predicted},i} - P_{\text{reported},i})^2}
    $$
    RMSE values provide the measure of magnitude error, directly comparable across epochs.
\end{itemize}

These metrics collectively capture both systematic (bias) and stochastic (sMAPE, RMSE) components of model error. The use of sMAPE complements MAPE-based assessments by reducing asymmetry in percentage errors, particularly important in low-population areas that dominate the rural extent of both case studies.

\subsubsection{Validation Strategy}

Validation was performed at two scales. First, at the \textit{district level}, aggregate accuracy metrics (Bias, sMAPE, RMSE) were computed to quantify improvements gained by introducing Hexagon imagery and local census data. Second, at the \textit{zonal level}, population totals were aggregated by \textit{GHS-SMOD} classes to assess how accurately each product reproduced functional urban structures. This dual approach enabled differentiation between overall model performance and spatial representativeness within urban, peri-urban, and rural contexts.

\subsection{Data and Software Availability}

All datasets used in this study are publicly available or derived from open-access sources. The \textit{GHS-SMOD} and \textit{GHS-POP} layers were obtained from the European Commission’s Joint Research Centre (JRC). Census data were digitised and harmonised from official publications of the Turkish Statistical Institute. A subset of the census data has been included in the appendix of this research (Table \ref{settlement_table}). The declassified Hexagon KH-9 imagery (1977) was accessed through the U.S. Geological Survey (USGS) EarthExplorer platform and are publicly available. 

All processing, including georeferencing, segmentation, and population disaggregation, was conducted in \textit{ArcGIS Pro} (version~3.5.4) using Python (ArcPy) for automation and open-source libraries (\texttt{pandas}, \texttt{rasterio}, \texttt{rioxarray}). Derived datasets and scripts used to reproduce the figures and analyses are available upon request or via the project’s GitHub repository\footnote{\href{https://github.com/pjgerrits/hexagon_grid_historical_pop.git}{}}.

\section{Results}

The results presented here evaluate three population grid products developed for the Arnavutköy and Çekmeköy areas of Istanbul, comparing their spatial patterns and aggregate performance across multiple scales. The objective was to assess whether Hexagon KH-9 (1977) built-up information could improve the internal accuracy of early-period population distributions relative to standard GHSL products. Each product maintains consistent total population counts but differs in how built-up surfaces are represented and used for spatial disaggregation: 

\begin{enumerate}
    \item \textbf{Baseline (GHS-POP 1975)} --- Derived from Landsat MSS-based \textit{GHS-BUILT-S} using the symbolic machine learning (SML) classification and temporal  back-projection method introduced by \cite{pesaresi_operating_2016}. The 1975 epoch is constrained by a monotonic assumption of spatial shrinkage of built-up areas over time that does not take into consideration demolition and reconstruction dynamics \cite{pesaresi_global_2016, taubenbock_monitoring_2012, freire_development_2016}. 
    \item \textbf{Hexagon Built-up (GHS-POP + Hexagon Built-up)} --- Population totals from GHS-POP are reallocated within the AOIs using built-up masks derived from semantically segmented Hexagon KH-9 imagery. 
    \item \textbf{Hexagon + Census (Census totals + Hexagon Built-up)} --- Local census counts (1935–2024) replace GHS population totals, redistributed using the same Hexagon built-up surface.
\end{enumerate}

Both districts represent peri-urban landscapes undergoing substantial transformation since the 1970s. The comparison highlights how the built-up proxy influences population disaggregation and evaluates consistency against historical census records and \textit{GHS- as SMOD} Degree of Urbanisation (DoU) classes.

\subsection{Pixel-wise comparison of population grid products}

\begin{figure}[ht!]
    \centering
    \includegraphics[width=\textwidth]{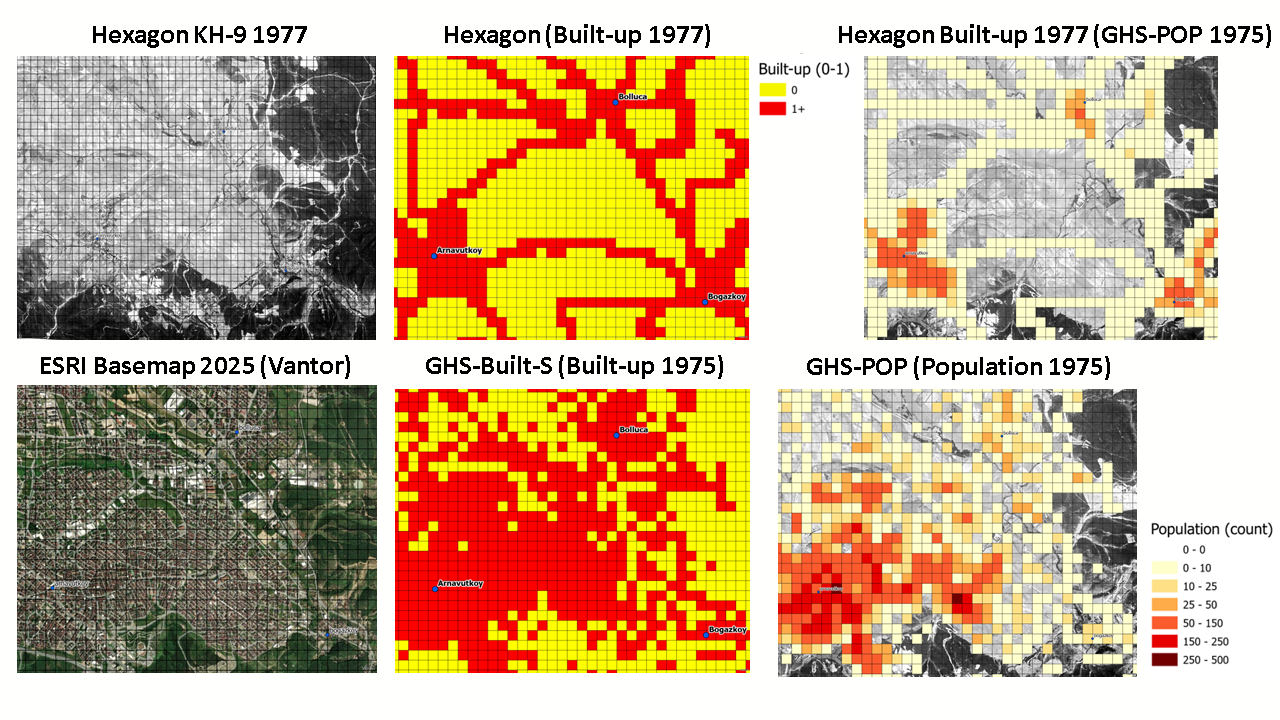}
    \caption{Comparison of built-up and population layers for Arnavutköy (1975--1977). 
    The Hexagon imagery (KH-9, 1977) provides a detailed reference of settlement morphology at the time. 
    The \textit{GHS-BUILT-S (1975)} layer, however, exhibits extensive built-up coverage overlapping with the 2025 basemap, indicating backcasting artefacts. 
    The \textit{GHS-POP (1975)} population surface mirrors this overextension, spreading population across rural terrain. 
    By contrast, the Hexagon-based weighting confines population to compact, road-connected built-up clusters visible in the historical imagery.}
    \label{fig:PopGrid_Builtup_Combined_Sample}
\end{figure}

Figure~\ref{fig:PopGrid_Builtup_Combined_Sample} illustrates how differences in the underlying built-up data affect population allocation.  
The \textit{GHS-BUILT-S (1975)} layer, derived from backcasting Landsat imagery through the Symbolic Machine Learning (SML) framework, enforces a monotonic temporal constraint that prevents contraction of built-up areas in earlier epochs \cite{pesaresi_advances_2024, freire_development_2016, leyk_spatial_2019}.  
Consequently, it includes suburban and infrastructural areas that were not yet present in the mid-1970s, leading to diffuse population allocation in \textit{GHS-POP (1975)}.  
In contrast, the Hexagon-derived built-up layer captures actual 1977 settlement morphology at sub-10~m resolution, providing a more temporally and spatially accurate weighting surface for dasymetric redistribution.  
The following pixel-wise analyses examine these effects for Arnavutköy and Çekmeköy separately.

\subsubsection*{Arnavutköy}

\begin{figure}[ht!]
    \centering
    \includegraphics[width=\textwidth]{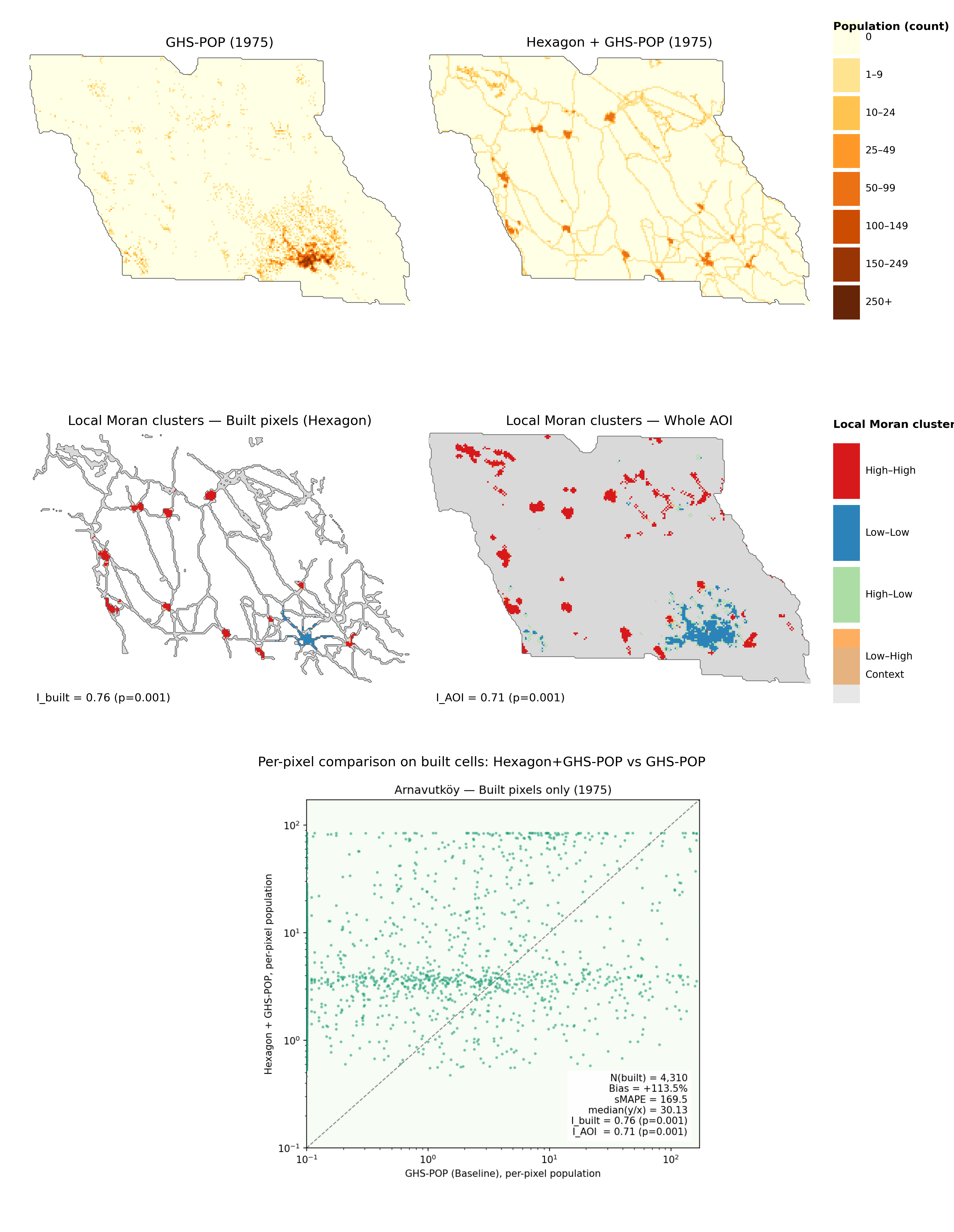}
    \caption{Pixel-wise comparison results for Arnavutköy (1975). 
    Top: population distribution for GHS-POP and Hexagon-weighted variants. 
    Middle: local Moran’s I clusters computed from per-pixel differences, for built-up pixels (left) and the full AOI (right). 
    Bottom: scatter plot of per-pixel population density in built-up cells. 
    High–High (red) and Low–Low (blue) clusters indicate spatially structured deviation patterns.}
    \label{fig:arnavutkoy_pixelwise}
\end{figure}

In Arnavutköy (Figure~\ref{fig:arnavutkoy_pixelwise}), the baseline \textit{GHS-POP (1975)} grid assigns population broadly across the southern plains and agricultural fields, while the Hexagon-weighted variant concentrates density within compact, linearly connected built-up clusters.  
These correspond closely to visible urban morphology in the 1977 imagery.  
The per-pixel scatter plot shows that Hexagon-weighted values are generally higher within built-up pixels, with a median(y/x) ratio of approximately 30.  
Spatial autocorrelation analysis supports this: the local Moran’s I values indicate strong clustering of residuals (\textit{I\textsubscript{built}} = 0.76, \textit{I\textsubscript{AOI}} = 0.71; $p < 0.01$).  
High–High clusters (red) identify dense cores underrepresented in GHS-POP, particularly near Arnavutköy’s historical town center and major road junctions, whereas Low–Low clusters (blue) correspond to peripheral rural zones that received excessive population allocation from the GHS-POP distribution.  
This pattern reveals a systematic improvement in the spatial coherence of population distribution when temporally synchronous built-up data are used.

\subsubsection*{Çekmeköy}

\begin{figure}[ht!]
    \centering
    \includegraphics[width=\textwidth]{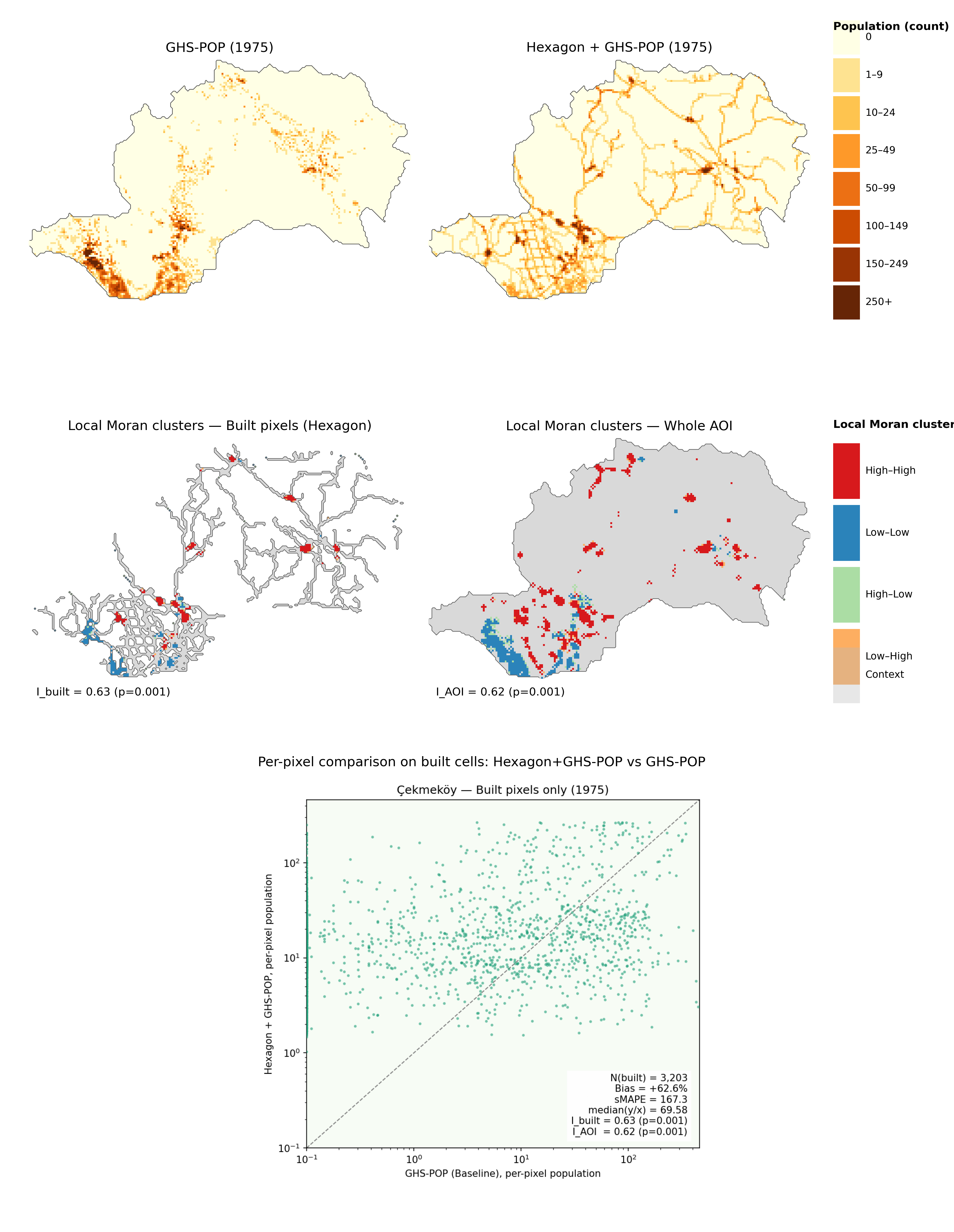}
    \caption{Pixel-wise comparison results for Çekmeköy (1975). 
    Top: population distribution for GHS-POP and Hexagon-weighted variants. 
    Middle: local Moran’s I clusters for built-up pixels (left) and the full AOI (right). 
    Bottom: scatter plot of per-pixel population density in built-up cells.}
    \label{fig:cekmekoy_pixelwise}
\end{figure}

For Çekmeköy (Figure~\ref{fig:cekmekoy_pixelwise}), the baseline \textit{GHS-POP (1975)} surface again exhibits widespread population allocation across unbuilt terrain, while the Hexagon-weighted output shows a much sharper distinction between built-up corridors and open areas.  
Although both datasets maintain identical total population sums between the baseline and hexagon-enhanced outputs, the internal distribution differs markedly: built-up pixels in the Hexagon-weighted grid contain roughly 7–8 times the population density of equivalent cells in the baseline product (median(y/x) = 7.8).  
Spatial clustering statistics are slightly lower than for Arnavutköy (\textit{I\textsubscript{built}} = 0.63, \textit{I\textsubscript{AOI}} = 0.62; $p < 0.01$), reflecting the less compact urban morphology of Çekmeköy in 1975.  
High–High clusters coincide with the historical settlement cores of Çekmeköy and Alemdağ, while Low–Low clusters dominate forested southern and northern fringes.  
These results indicate that the Hexagon-based redistribution more accurately reflects the discontinuous and corridor-like development characteristic of Çekmeköy’s 1970s settlement pattern.

\subsubsection*{Synthesis and implications}

\begin{figure}[ht!]
    \centering
    \includegraphics[width=1\linewidth]{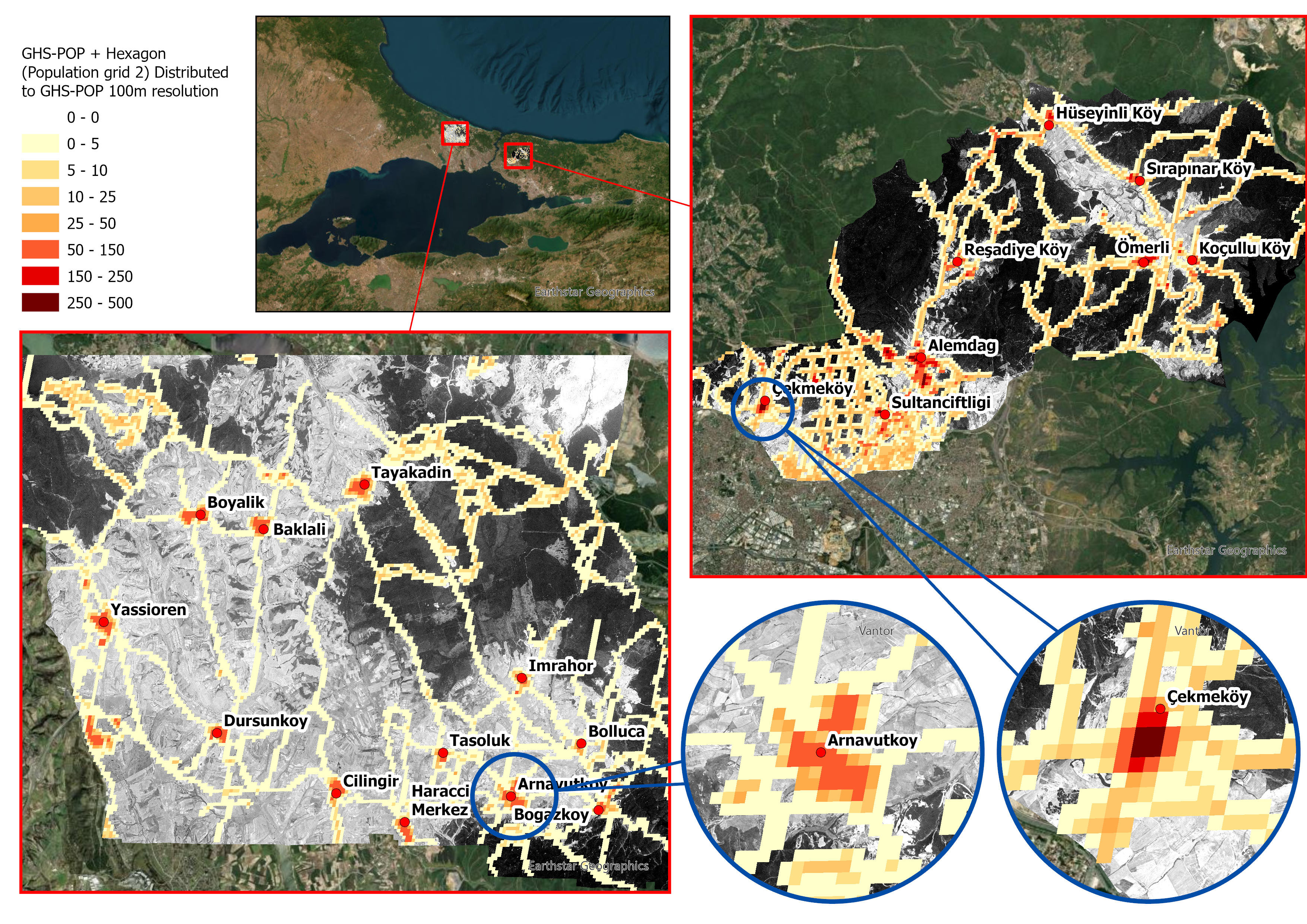}
    \caption{Comparison of population distributions for Arnavutköy and Çekmeköy. The Hexagon KH-9 derived built-up mask captures linear and smaller settlement structures omitted in the GHSL product better.}
    \label{fig:arnavutkoy_hexagon_population_comparison}
\end{figure}

Across both case studies, the pixel-wise comparison highlights that spatial accuracy in early-epoch population grids is strongly dependent on the temporal accuracy of the built-up input.  
Even if we do not consider \textit{GHS-POP (1975)} overestimation of total administrative totals for the region, its coarse, backcasted built-up surfaces distribute population unrealistically across rural land in our study area. Replacing these with contemporaneous, high-resolution Hexagon data improves spatial distribution and also introduces detectable structure in the residuals, evidenced by the significant Moran’s I values. This confirms that spatial error in the baseline product is not random but systematically aligned with built-up misclassification. Our hexagon-based approach therefore yields a coherent and historically more plausible spatial representation of population in the area as can be seen on figure~\ref{fig:arnavutkoy_hexagon_population_comparison}.

\subsection{Zonal Comparison by GHS-SMOD Degree of Urbanisation Classes}

\begin{table}
\caption{Population counts by GHS-SMOD class in 1975 for each product.}
\label{tab:smod_pop_1975}
\begin{tabular}{llrrr}
\toprule
AOI & SMOD class & GHS-POP & Hexagon + GHS-POP & Census + Hexagon\\
\midrule
Arnavutköy & Urban centre & Not present& Not present& Not present\\
Arnavutköy & Dense urban cluster & 20{,}508 & 3{,}008 & 546 \\
Arnavutköy & Suburban / peri-urban & 9{,}179 & 4{,}157 & 755 \\
Arnavutköy & Low-density rural & 5{,}270 & 12{,}084 & 2{,}195 \\
Arnavutköy & Very low-density rural & 2{,}135 & 17{,}709 & 3{,}217 \\
Çekmeköy & Urban centre & 60{,}814 & 26{,}126 & 2{,}811 \\
Çekmeköy & Dense urban cluster & Not present& Not present& Not present\\
Çekmeköy & Suburban / peri-urban & 2{,}380 & 5{,}562 & 598 \\
Çekmeköy & Low-density rural & 5{,}148 & 18{,}905 & 2{,}034 \\
Çekmeköy & Very low-density rural & 565 & 14{,}078 & 1{,}515 \\
\bottomrule
\end{tabular}
\end{table}

\begin{figure}[h!]
\centering
\includegraphics[width=\linewidth]{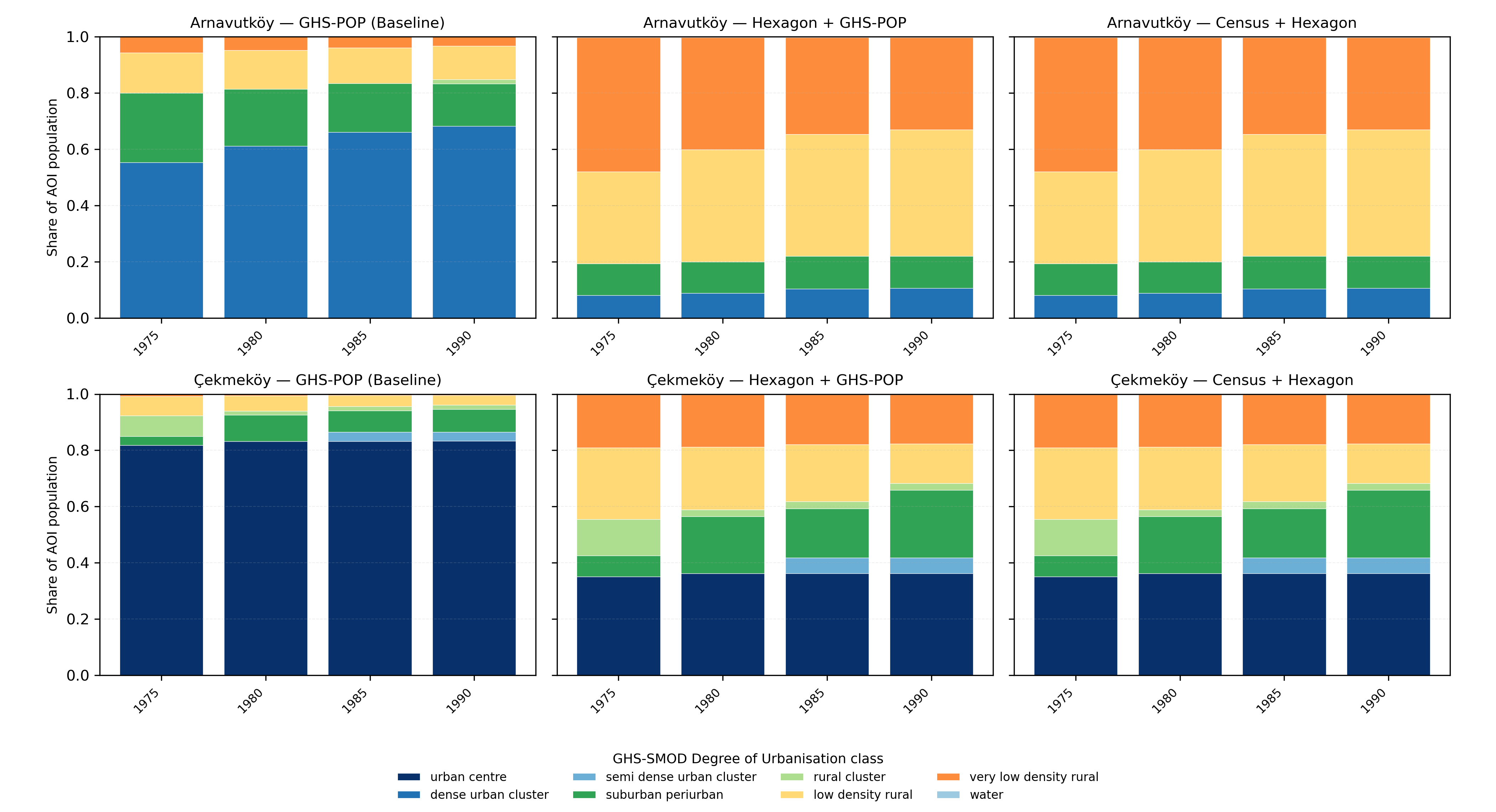}
\caption{Relative population share by GHS-SMOD Degree of Urbanisation class (1975--1990). Each bar represents the total population of an AOI and year, normalised to 100\%.}
\label{fig:smod_pop_share}
\end{figure}

To assess how the inclusion of Hexagon-built surfaces modifies the spatial allocation of population across functional urbanisation categories, all products were aggregated to \textit{GHS-SMOD} Degree of Urbanisation (DoU) classes for the four target years (1975–1990). The results reveal significant structural differences between the baseline \textit{GHS-POP} and the Hexagon-enhanced grids (Table~\ref{tab:smod_pop_1975}, Figure~\ref{fig:smod_pop_share}).

In both Arnavutköy and Çekmeköy, the \textit{GHS-POP 1975} baseline concentrates more than two-thirds of the total population within ``urban centre'' or ``dense urban cluster'' categories. This over-representation of urban classes stems from the coarse spatial delineation of the \textit{GHS-BUILT-S (1975)} layer, which captures extensive areas of continuous settlement that, in 1975, were largely non-built or agricultural. The backcasting of built-up surfaces from more recent imagery effectively inflates the apparent extent of dense settlement for early epochs.

By contrast, the \textit{Hexagon + GHS-POP} and \textit{Census + Hexagon} products redistribute a substantially greater share of the population into ``suburban/peri-urban'' and ``low-density rural'' classes—categories that more accurately reflect the dispersed and transitional character of northern Istanbul’s urban fringe during the 1970s. In Arnavutköy, the share of population assigned to rural and peri-urban SMOD classes rises from roughly 30\% in the baseline to over 80\% in the Hexagon-enhanced grid. Similarly, in Çekmeköy, the proportion of population allocated to dense or urban classes decreases from about 85\% in the baseline to less than 50\% once Hexagon data are introduced. These shifts illustrate that the refined built-up mask suppresses over-allocation to high-density zones and restores a more realistic population gradient along the rural–urban continuum.

This zonal pattern aligns with the findings of the pixel-wise comparison results, which showed significant positive spatial autocorrelation ($I_{\text{built}} = 0.76$–$0.63$, $p = 0.001$) between the Hexagon-based and baseline distributions, but a systematic bias toward denser cells in the latter. The DoU aggregation results thus provide complementary evidence that the GHS-POP 1975 surface systematically overestimates the spatial extent and intensity of urbanisation.

Overall, the inclusion of Hexagon KH-9 built-up data improves the consistency between population distribution and the morphological reality of early-period settlement patterns. The resulting population surfaces more faithfully represent the functional hierarchy of urban, peri-urban, and rural zones while maintaining census total constraints. These improvements strengthen the validity of the Hexagon-enhanced population grids as temporally and spatially coherent representations of mid-20th-century urban structure in Istanbul.

\section{Discussion}

The results of this study provide evidence for the utility of declassified high-resolution reconnaissance imagery in reconstructing historical population distributions. By integrating Hexagon KH-9 data with harmonised census records, the proposed workflow corrects the systematic overestimation and spatial bias characteristic of global population datasets such as GHSL and HYDE. In doing so, it directly addresses long-standing concerns regarding the reliability of global gridded products in heterogeneous peri-urban and rural contexts \cite{lang-ritter_global_2025, leyk_spatial_2019}.

While Hexagon imagery has previously found applications in disciplines such as archaeology, environmental change, and glacier monitoring, its use for quantitative urban and demographic reconstruction remains limited \cite{kabadayi_agricultural_2022}. Earlier studies employing Key Hole  (KH) data have largely relied on visual interpretation or manual vectorisation, both of which constrain scalability and introduce subjectivity. In contrast, our study demonstrates a semi-automated, reproducible method for extracting built-up surfaces from declassified satellite imagery, capable of supporting consistent, quantitative population modelling.

\subsection{Implications for Historical Population Modelling}

The Istanbul case studies highlight several broader implications.  
First, \textbf{spatial resolution} is decisive for accurate population disaggregation. The sub-metre built-up mask derived from Hexagon imagery captures dispersed and transitional settlement forms that were systematically omitted in the coarser, back-cast \textit{GHS-BUILT-S (1975)} layer. As a result, the baseline \textit{GHS-POP} product allocates an excessive share of population to dense urban SMOD classes, often exceeding two-thirds of the total population in 1975, despite limited physical development evident in the corresponding historical imagery. The Hexagon-enhanced variants, by contrast, shift the population distribution toward suburban and rural classes—over 80\% in Arnavutköy and around 50\% in Çekmeköy—aligning far more closely with the morphological reality of the 1970s fringe landscape. These shifts confirm that the back-casting of built-up surfaces from later periods introduces temporal bias, which can be substantially reduced by incorporating synchronous high-resolution data.

Second, \textbf{the integration of legacy imagery and census data} enables a temporally consistent reconstruction of demographic patterns at a spatial resolution previously unavailable for the mid-20th century. The Hexagon KH-9 archive effectively bridges the data gap between early Landsat imagery (post-1972) and historical censuses, permitting a reliable reconstruction of pre-satellite urbanisation dynamics.

Finally, \textbf{the methodological transferability} of this workflow extends beyond Istanbul. The pipeline developed here, combining precise georeferencing, built-up classification, and dasymetric population redistribution, can be adapted to other regions where declassified or archival high-resolution imagery is available. This adaptability is particularly relevant for areas where traditional remote-sensing time series fail to capture early urban expansion or where historical census data remain underutilised.

\subsection{Reassessing the Fitness for Use of Global Products}

These findings underscore that the “fitness for use” of global population datasets is inherently scale-, period- and context-dependent. While global frameworks such as GHSL and HYDE are indispensable for long-term and cross-national analyses, their early-epoch layers remain constrained in accuracy for fine-scale applications such as urban morphology studies, vulnerability assessments, and climate adaptation planning. The hexagon-enhanced grids demonstrate that incorporating temporally synchronous, high-resolution data can substantially improve both the morphological and functional representation of historical population surfaces. In this sense, the approach presented here complements recent calls for hybrid, multi-source reconstruction strategies that integrate global models with local evidence and GeoAI-based enhancement \cite{uhl_towards_2025, sertel_hexalcseg_2024}.

\subsection{Limitations and Future Work}
While our study demonstrates substantial advances, several limitations working with historical data should be acknowledged. 

First, although Hexagon KH-9 imagery offers uniquely high spatial resolution for its era, its global coverage is not uniform, and its temporal availability is restricted to the 1970s and 1980s. Furthermore, image quality can vary due to atmospheric conditions, sensor artefacts, or film scanning issues, which can impact classification accuracy.

Second, the processing of this legacy imagery remains a non-trivial task. The limited spectral (panchromatic) and radiometric resolution, combined with significant geometric distortions, presents challenges for automated classification. While our workflow mitigates these issues, uncertainties can be introduced during georeferencing and segmentation, potentially propagating errors into the final population grids.

Third, the historical census data, though unusually granular for our study area, are not immune to reporting inconsistencies or changes in administrative boundaries over time. The process of harmonising historical records to contemporary boundaries inevitably introduces a degree of spatial uncertainty, which may affect error metrics.

Future research should focus on addressing these limitations. Expanding the archival data sources to include other declassified assets like CORONA imagery or historical aerial photography could extend the temporal and geographic applicability of this approach. Methodologically, the use of advanced machine learning techniques, such as semi-supervised or transfer learning models, could improve segmentation performance across diverse geographies with less manual intervention on a larger scale for built-up spaces \cite{schug_quantifying_2025}. Finally, developing probabilistic approaches to population disaggregation could help to formally quantify and propagate the uncertainties from both the imagery and the census inputs into the final gridded estimates.

\section{Conclusion}
This study illustrates that the incorporation of semantically segmented Hexagon imagery with standardised, geocoded census data improves the spatial and temporal accuracy of historical population grids in Istanbul's peri-urban areas. Through the comparison of a GHSL-style baseline, a Hexagon-enhanced variant, and a Hexagon+local census product, we demonstrate that high-resolution legacy imagery mitigates systematic overestimation and spatial misallocation, while localised census inputs further refine intra-district allocations and growth trajectories. This methodology is applicable in various contexts with appropriate archival imagery and census data, providing a pragmatic avenue to enhance and deepen the historical context of global gridded population products for long-term applications \cite{leyk_spatial_2019,schiavina_ghs-pop_2023}.

Enhanced historical population grids with fine spatial resolutions facilitate more informed evaluations of long-term urbanisation trends, including the detection of multi-decadal growth hotspots and the measurement of urban expansion rates traceable to the 1970s \cite{schug_quantifying_2025,corbane_automated_2019}. These grids provide essential input data for the calibration and validation of urban growth models, land-use change models, and integrated assessment frameworks that necessitate consistent, spatially explicit population trajectories spanning several decades \cite{uhl_combining_2021,uhl_towards_2025}. Concurrently, enhanced historical population distributions provide a more robust empirical basis for retrospective analyses in climate research, enabling researchers to reconstruct previous settlement patterns and densities that affect local climate conditions and vulnerability evaluations \cite{bechtel_beyond_2018}. This methodology can be applied to other rapidly urbanising areas where Hexagon imagery and historical census data exist, enhancing the understanding of global urbanisation dynamics and their enduring socio-environmental consequences.

\section*{Acknowledgments}
Support details have been anonymised for peer review.

\section*{Funding details}
This work was supported by the European
Research Council under Grant IDs: 679097 (UrbanOccupationsOETR Project) and 679097 (GeoAI LULC Seg), PI: M. Erdem Kabadayi; and UK Research and Innovation Future Leaders Fellowships under Grant References MR/Y011856/1 and MR/S01795X/2, PI: Ana Basiri.

\section*{Declaration of Interests}
The authors declare no competing interests.

\section*{Data Availability Statement}
Census data are publicly available from the Turkish Statistical Institute. GHSL data are available at the EU-JRC GHSL portal. Additional census information has been included as supplementary material in the appendix, and additional data are available upon request.

\printbibliography

\appendix
\break
\section{Appendix}

\begin{sidewaystable}[p]
\centering
\caption{Historical AoI census totals, growth rates, GHS-POP estimates, and percentage differences.}
\setlength{\tabcolsep}{4pt}
\scriptsize
\resizebox{\textheight}{!}{%
\begin{tabular}{l rr r r r r r rr rr}
\toprule
 &  &  &  & \multicolumn{4}{c}{TÜİK Census data} & \multicolumn{2}{c}{GHS-POP estimates} & \multicolumn{2}{c}{Percentage Difference} \\
\cmidrule(lr){5-8}\cmidrule(lr){9-10}\cmidrule(lr){11-12}
year & Arnavutköy & Çekmeköy & Grand Total & Arnavutköy Growth Rate & Arnavutköy Yearly Growth Rate & Çekmeköy Growth Rate  & Çekmeköy Yearly Growth Rate & Arnavutköy & Cekmekoy & arn\_percent\_diff & cek\_percent\_diff \\
\midrule
1935 & 5662 & 2551 & 8213 &  &  &  &  &  &  &  &  \\
1955 & 9613 & 6477 & 16090 & 0.697809961 & 0.026820319 & 1.539004312 & 0.0476909 &  &  &  &  \\
1970 & 8308 & 3889 & 12197 & -0.135753667 & -0.009679347 & -0.399567701 & -0.033435286 &  &  &  &  \\
1975 & 8427 & 8018 & 16445 & 0.014323544 & 0.002848435 & 1.061712522 & 0.155701353 & 37105.65552 & 74876.01411 & 23\% & 11\% \\
1980 & 11187 & 10926 & 22113 & 0.32751869 & 0.058298374 & 0.362683961 & 0.063846646 & 43540.08832 & 88662.9142 & 26\% & 12\% \\
1985 & 14659 & 16420 & 31079 & 0.31036024 & 0.055548375 & 0.502837269 & 0.08488172 & 51544.28422 & 105794.1459 & 28\% & 16\% \\
1990 & 42150 & 34420 & 76570 & 1.875366669 & 0.235203879 & 1.096224117 & 0.159544827 & 60526.97139 & 125138.1648 & 70\% & 28\% \\
2000 & 96528 & 88083 & 184611 & 1.290106762 & 0.086389534 & 1.559064497 & 0.098520392 & 81038.85455 & 178118.0686 & 119\% & 49\% \\
2007 & 125612 & 139898 & 265510 & 0.301301177 & 0.038340249 & 0.58825199 & 0.068323478 & 100754.3684 & 221359.4825 & 125\% & 63\% \\
2008 & 129830 & 133750 & 263580 & 0.033579594 & 0.033579594 & -0.043946304 & -0.043946304 &  &  &  &  \\
2009 & 139211 & 140551 & 279762 & 0.072256027 & 0.072256027 & 0.050848598 & 0.050848598 &  &  &  &  \\
2010 & 145911 & 152420 & 298331 & 0.048128381 & 0.048128381 & 0.084446215 & 0.084446215 & 123233.875 & 269882.798 & 118\% & 56\% \\
2011 & 151774 & 165433 & 317207 & 0.040182029 & 0.040182029 & 0.085375935 & 0.085375935 &  &  &  &  \\
2012 & 157240 & 174080 & 331320 & 0.036014074 & 0.036014074 & 0.052268894 & 0.052268894 &  &  &  &  \\
2013 & 164412 & 186281 & 350693 & 0.045611804 & 0.045611804 & 0.070088465 & 0.070088465 &  &  &  &  \\
2014 & 172282 & 198461 & 370743 & 0.047867552 & 0.047867552 & 0.06538509 & 0.06538509 &  &  &  &  \\
2015 & 181335 & 208366 & 389701 & 0.052547567 & 0.052547567 & 0.04990905 & 0.04990905 & 132321.3618 & 292591.751 & 137\% & 71\% \\
2016 & 190930 & 215454 & 406384 & 0.052913117 & 0.052913117 & 0.034017066 & 0.034017066 &  &  &  &  \\
2017 & 201961 & 223807 & 425768 & 0.057775101 & 0.057775101 & 0.038769296 & 0.038769296 &  &  &  &  \\
2018 & 207614 & 227355 & 434969 & 0.027990553 & 0.027990553 & 0.015852945 & 0.015852945 &  &  &  &  \\
2019 & 216282 & 239086 & 455368 & 0.041750556 & 0.041750556 & 0.051597722 & 0.051597722 &  &  &  &  \\
2020 & 227090 & 248133 & 475223 & 0.049971796 & 0.049971796 & 0.03783994 & 0.03783994 & 130520.9175 & 298978.7085 & 174\% & 83\% \\
2021 & 238864 & 262591 & 501455 & 0.051847285 & 0.051847285 & 0.058267139 & 0.058267139 &  &  &  &  \\
2022 & 251347 & 270185 & 521532 & 0.052259863 & 0.052259863 & 0.028919498 & 0.028919498 &  &  &  &  \\
2023 & 259455 & 274735 & 534190 & 0.032258193 & 0.032258193 & 0.016840313 & 0.016840313 &  &  &  &  \\
2024 & 266306 & 281320 & 547626 & 0.02640535 & 0.02640535 & 0.023968552 & 0.023968552 &  &  &  &  \\
\bottomrule
\label{growth_table}
\end{tabular}%
}
\end{sidewaystable}

\begin{table}[p]
\centering
\tiny 
\caption{Temporal population and location data for settlements in AoIs (LAU-2 level).}
\begin{tabular}{|l|l|l|l|l|l|l|l|}
\hline
\textbf{AoI} & \textbf{Province} & \textbf{District} & \textbf{Settlement} & \textbf{Pop Count} & \textbf{year} & \textbf{x} & \textbf{y} \\ \hline
Arnavutköy   & İstanbul          & Çatalca           & Baklalı             & 375            & 1975          & 28.65      & 41.26      \\ \hline
Arnavutköy   & İstanbul          & Çatalca           & Baklalı             & 368            & 1980          & 28.65      & 41.26      \\ \hline
Arnavutköy   & İstanbul          & Çatalca           & Baklalı             & 407            & 1985          & 28.65      & 41.26      \\ \hline
Arnavutköy   & İstanbul          & Çatalca           & Baklalı             & 586            & 1990          & 28.65      & 41.26      \\ \hline
Arnavutköy   & İstanbul          & Çatalca           & Boyalık             & 456            & 1975          & 28.63      & 41.26      \\ \hline
Arnavutköy   & İstanbul          & Çatalca           & Boyalık             & 676            & 1980          & 28.63      & 41.26      \\ \hline
Arnavutköy   & İstanbul          & Çatalca           & Boyalık             & 495            & 1985          & 28.63      & 41.26      \\ \hline
Arnavutköy   & İstanbul          & Çatalca           & Boyalık             & 485            & 1990          & 28.63      & 41.26      \\ \hline
Arnavutköy   & İstanbul          & Çatalca           & Dursun              & 392            & 1975          & 28.64      & 41.20      \\ \hline
Arnavutköy   & İstanbul          & Çatalca           & Dursunköy           & 543            & 1980          & 28.64      & 41.20      \\ \hline
Arnavutköy   & İstanbul          & Çatalca           & Dursunköy           & 558            & 1985          & 28.64      & 41.20      \\ \hline
Arnavutköy   & İstanbul          & Çatalca           & Dursunköy           & 635            & 1990          & 28.64      & 41.20      \\ \hline
Arnavutköy   & İstanbul          & Çatalca           & Durusu              & 2213           & 1985          & 28.68      & 41.30      \\ \hline
Arnavutköy   & İstanbul          & Çatalca           & Durusu              & 3018           & 1990          & 28.68      & 41.30      \\ \hline
Arnavutköy   & İstanbul          & Çatalca           & Durusu (Ptt)        & 1688           & 1975          & 28.68      & 41.30      \\ \hline
Arnavutköy   & İstanbul          & Çatalca           & Durusu (Ptt)        & 1995           & 1980          & 28.68      & 41.30      \\ \hline
Arnavutköy   & İstanbul          & Çatalca           & Yassıören           & 561            & 1975          & 28.60      & 41.23      \\ \hline
Arnavutköy   & İstanbul          & Çatalca           & Yassıören           & 557            & 1980          & 28.60      & 41.23      \\ \hline
Arnavutköy   & İstanbul          & Çatalca           & Yassıören           & 555            & 1985          & 28.60      & 41.23      \\ \hline
Arnavutköy   & İstanbul          & Çatalca           & Yassıören           & 609            & 1990          & 28.60      & 41.23      \\ \hline
Arnavutköy   & İstanbul          & Gaziosmanpaşa     & Arnavutköy          & 4182           & 1985          & 28.74      & 41.18      \\ \hline
Arnavutköy   & İstanbul          & Gaziosmanpaşa     & Arnavutköy          & 21143          & 1990          & 28.74      & 41.18      \\ \hline
Arnavutköy   & İstanbul          & Gaziosmanpaşa     & Arnavutköy (Ptt)    & 1551           & 1975          & 28.74      & 41.18      \\ \hline
Arnavutköy   & İstanbul          & Gaziosmanpaşa     & Arnavutköy (Ptt)    & 2221           & 1980          & 28.74      & 41.18      \\ \hline
Arnavutköy   & İstanbul          & Gaziosmanpaşa     & Boğazköy            & 755            & 1975          & 28.77      & 41.18      \\ \hline
Arnavutköy   & İstanbul          & Gaziosmanpaşa     & Boğazköy            & 963            & 1980          & 28.77      & 41.18      \\ \hline
Arnavutköy   & İstanbul          & Gaziosmanpaşa     & Boğazköy            & 1249           & 1985          & 28.77      & 41.18      \\ \hline
Arnavutköy   & İstanbul          & Gaziosmanpaşa     & Boğazköy            & 4495           & 1990          & 28.77      & 41.18      \\ \hline
Arnavutköy   & İstanbul          & Gaziosmanpaşa     & Bolluca             & 314            & 1975          & 28.77      & 41.20      \\ \hline
Arnavutköy   & İstanbul          & Gaziosmanpaşa     & Bolluca             & 777            & 1980          & 28.77      & 41.20      \\ \hline
Arnavutköy   & İstanbul          & Gaziosmanpaşa     & Bolluca             & 913            & 1985          & 28.77      & 41.20      \\ \hline
Arnavutköy   & İstanbul          & Gaziosmanpaşa     & Bolluca             & 2409           & 1990          & 28.77      & 41.20      \\ \hline
Arnavutköy   & İstanbul          & Gaziosmanpaşa     & Çilingir            & 491            & 1975          & 28.68      & 41.19      \\ \hline
Arnavutköy   & İstanbul          & Gaziosmanpaşa     & Çilingir            & 520            & 1980          & 28.68      & 41.19      \\ \hline
Arnavutköy   & İstanbul          & Gaziosmanpaşa     & Çilingir            & 550            & 1985          & 28.68      & 41.19      \\ \hline
Arnavutköy   & İstanbul          & Gaziosmanpaşa     & Çilingir            & 612            & 1990          & 28.68      & 41.19      \\ \hline
Arnavutköy   & İstanbul          & Gaziosmanpaşa     & Haraççı             & 484            & 1975          & 28.70      & 41.18      \\ \hline
Arnavutköy   & İstanbul          & Gaziosmanpaşa     & Haraççı             & 745            & 1980          & 28.70      & 41.18      \\ \hline
Arnavutköy   & İstanbul          & Gaziosmanpaşa     & Haraççı             & 897            & 1985          & 28.70      & 41.18      \\ \hline
Arnavutköy   & İstanbul          & Gaziosmanpaşa     & Haraççı             & 2671           & 1990          & 28.70      & 41.18      \\ \hline
Arnavutköy   & İstanbul          & Gaziosmanpaşa     & İmrahor             & 270            & 1975          & 28.75      & 41.22      \\ \hline
Arnavutköy   & İstanbul          & Gaziosmanpaşa     & İmrahor             & 555            & 1980          & 28.75      & 41.22      \\ \hline
Arnavutköy   & İstanbul          & Gaziosmanpaşa     & İmrahor             & 958            & 1985          & 28.75      & 41.22      \\ \hline
Arnavutköy   & İstanbul          & Gaziosmanpaşa     & İmrahor             & 1863           & 1990          & 28.75      & 41.22      \\ \hline
Arnavutköy   & İstanbul          & Gaziosmanpaşa     & Taşoluk             & 505            & 1975          & 28.72      & 41.20      \\ \hline
Arnavutköy   & İstanbul          & Gaziosmanpaşa     & Taşoluk             & 589            & 1980          & 28.72      & 41.20      \\ \hline
Arnavutköy   & İstanbul          & Gaziosmanpaşa     & Taşoluk             & 747            & 1985          & 28.72      & 41.20      \\ \hline
Arnavutköy   & İstanbul          & Gaziosmanpaşa     & Taşoluk             & 2527           & 1990          & 28.72      & 41.20      \\ \hline
Arnavutköy   & İstanbul          & Gaziosmanpaşa     & Tayakadın           & 585            & 1975          & 28.69      & 41.27      \\ \hline
Arnavutköy   & İstanbul          & Gaziosmanpaşa     & Tayakadın           & 678            & 1980          & 28.69      & 41.27      \\ \hline
Arnavutköy   & İstanbul          & Gaziosmanpaşa     & Tayakadın           & 935            & 1985          & 28.69      & 41.27      \\ \hline
Arnavutköy   & İstanbul          & Gaziosmanpaşa     & Tayakadın           & 1097           & 1990          & 28.69      & 41.27      \\ \hline
Çekmeköy     & İstanbul          & Beykoz            & Koçullu             & 210            & 1975          & 29.35      & 41.08      \\ \hline
Çekmeköy     & İstanbul          & Beykoz            & Koçullu             & 292            & 1980          & 29.35      & 41.08      \\ \hline
Çekmeköy     & İstanbul          & Beykoz            & Koçullu             & 326            & 1985          & 29.35      & 41.08      \\ \hline
Çekmeköy     & İstanbul          & Beykoz            & Muratlı             & 0              & 1975          & 29.35      & 41.05      \\ \hline
Çekmeköy     & İstanbul          & Beykoz            & Muratlı             & 0              & 1980          & 29.35      & 41.05      \\ \hline
Çekmeköy     & İstanbul          & Beykoz            & Ömerli (Bm)         & 1832           & 1985          & 29.33      & 41.08      \\ \hline
Çekmeköy     & İstanbul          & Beykoz            & Ömerli (Bm)         & 2177           & 1990          & 29.33      & 41.08      \\ \hline
Çekmeköy     & İstanbul          & Beykoz            & Ömerli (Ptt)        & 1090           & 1975          & 29.33      & 41.08      \\ \hline
Çekmeköy     & İstanbul          & Beykoz            & Ömerli (Ptt)        & 1472           & 1980          & 29.33      & 41.08      \\ \hline
Çekmeköy     & İstanbul          & Beykoz            & Sırapınar           & 247            & 1975          & 29.33      & 41.10      \\ \hline
Çekmeköy     & İstanbul          & Beykoz            & Sırapınar           & 323            & 1980          & 29.33      & 41.10      \\ \hline
Çekmeköy     & İstanbul          & Beykoz            & Sırapınar           & 376            & 1985          & 29.33      & 41.10      \\ \hline
Çekmeköy     & İstanbul          & Ümraniye          & Alemdar             & 3056           & 1975          & 29.24      & 41.05      \\ \hline
Çekmeköy     & İstanbul          & Ümraniye          & Alemdar             & 4044           & 1980          & 29.24      & 41.05      \\ \hline
Çekmeköy     & İstanbul          & Ümraniye          & Alemdar             & 5433           & 1985          & 29.24      & 41.05      \\ \hline
Çekmeköy     & İstanbul          & Ümraniye          & Alemdar             & 6684           & 1990          & 29.24      & 41.05      \\ \hline
Çekmeköy     & İstanbul          & Ümraniye          & Çekme               & 1850           & 1975          & 29.18      & 41.04      \\ \hline
Çekmeköy     & İstanbul          & Ümraniye          & Çekme               & 1938           & 1980          & 29.18      & 41.04      \\ \hline
Çekmeköy     & İstanbul          & Ümraniye          & Çekme               & 3789           & 1985          & 29.18      & 41.04      \\ \hline
Çekmeköy     & İstanbul          & Ümraniye          & Çekme               & 13523          & 1990          & 29.18      & 41.04      \\ \hline
Çekmeköy     & İstanbul          & Ümraniye          & Hüseyinli           & 340            & 1975          & 29.29      & 41.12      \\ \hline
Çekmeköy     & İstanbul          & Ümraniye          & Hüseyinli           & 402            & 1980          & 29.29      & 41.12      \\ \hline
Çekmeköy     & İstanbul          & Ümraniye          & Hüseyinli           & 448            & 1985          & 29.29      & 41.12      \\ \hline
Çekmeköy     & İstanbul          & Ümraniye          & Hüseyinli           & 586            & 1990          & 29.29      & 41.12      \\ \hline
Çekmeköy     & İstanbul          & Ümraniye          & Koçullu             & 485            & 1990          & 29.35      & 41.08      \\ \hline
Çekmeköy     & İstanbul          & Ümraniye          & Reşadiye            & 300            & 1975          & 29.25      & 41.08      \\ \hline
Çekmeköy     & İstanbul          & Ümraniye          & Reşadiye            & 420            & 1980          & 29.25      & 41.08      \\ \hline
Çekmeköy     & İstanbul          & Ümraniye          & Reşadiye            & 439            & 1985          & 29.25      & 41.08      \\ \hline
Çekmeköy     & İstanbul          & Ümraniye          & Reşadiye            & 778            & 1990          & 29.25      & 41.08      \\ \hline
Çekmeköy     & İstanbul          & Ümraniye          & Sırapınar           & 440            & 1990          & 29.33      & 41.10      \\ \hline
Çekmeköy     & İstanbul          & Ümraniye          & Sultançiftliği      & 925            & 1975          & 29.22      & 41.03      \\ \hline
Çekmeköy     & İstanbul          & Ümraniye          & Sultançiftliği      & 2035           & 1980          & 29.22      & 41.03      \\ \hline
Çekmeköy     & İstanbul          & Ümraniye          & Sultançiftliği      & 3777           & 1985          & 29.22      & 41.03      \\ \hline
Çekmeköy     & İstanbul          & Ümraniye          & Sultançiftliği      & 9747           & 1990          & 29.22      & 41.03      \\ \hline
\label{settlement_table}
\end{tabular}
\end{table}

\end{document}